\documentclass[a4paper,noindent]{JHEP3}
\usepackage{feynarts}
\usepackage{amsmath}
\usepackage{multirow}
\usepackage{mathenv}
\usepackage{array}
\usepackage{cite}
\usepackage{color}
\usepackage{slashed}
\let\normalcolor\relax

\newcommand{\bea}{\begin{eqnarray}}
\newcommand{\eea}{\end{eqnarray}}
\newcommand{\bean}{\begin{eqnarray*}}
\newcommand{\eean}{\end{eqnarray*}}

\def\bra#1{\left\langle #1\right|}
\def\sqbra#1{\left[ #1\right|}
\def\ket#1{\left| #1\right\rangle}
\def\sqket#1{\left| #1\right]}
\def\gb #1{ \left\langle #1 \right]}
\def\tgb #1{ \left[ #1 \right\rangle}
\def\cb #1{ \left[ #1 \right]}

\def\eps{\epsilon}

\def\vev#1{\left\langle #1 \right\rangle}

\def\ord{{\cal O}}

\def\Label#1{\label{#1} \smash{\hbox to0pt{\raise1ex\hbox{\tiny[#1]}\hss}}}

\newcommand{\be}{\begin{equation}}
\newcommand{\ee}{\end{equation}}
\newcommand{\bd}{\begin{displaymath}}
\newcommand{\ed}{\end{displaymath}}

\newcommand{\Spapb}[3]{\langle  {#1} | {#2} | {#3} ]}

\title{External leg corrections  in the unitarity method}
\author{Ruth Britto, Edoardo Mirabella\\
Institut de Physique Th\'eorique, CEA-Saclay, F-91191, Gif-sur-Yvette
cedex, France \\
Email: \email{ruth.britto@cea.fr,  edoardo.mirabella@cea.fr}}

\abstract{ 
Unitarity cuts diverge in the channel of a single massive external fermion.  We propose an off-shell continuation of the momentum that allows a finite evaluation of the unitarity cuts.  If the cut is taken with complete amplitudes on each side, our continuation and expansion around the on-shell configuration produces the finite contribution to the bubble coefficient.  Finite parts in the expansion of the external leg counterterms must be included explicitly as well.
}

\keywords{QCD, Feynman integrals, NLO calculations}
\preprint{IPhT-t11/185}

\begin{document}

\section{Introduction}

Precision computations beyond the leading order (LO)  play a crucial role in the physical analyses 
performed at the LHC. Several new-physics signals, as well as many background processes, 
involve multi-particle final states. At LO, these processes exhibit  a large scale 
uncertainty which can be softened by  including the next-to-leading order (NLO) contributions, in which one-loop amplitudes have been a challenging component in processes of current importance~\cite{Buttar:2006zd,Bern:2008ef}.

Recent years have seen rapid progress in computing one-loop amplitudes, due largely to the use of 
on-shell methods grouped under the name of ``generalized unitarity.''    One advantage of on-shell 
methods is that they enable the computation of loop amplitudes in terms of tree-level amplitudes, 
which are relatively easier to generate either numerically or analytically in compact forms.

Unitarity methods for one-loop amplitudes \cite{Cutkosky:1960sp,vanNeerven:1985xr,Bern:1994zx,Bern:1994cg,Bern:1997sc,Bern:2004ky,Britto:2004nc,Mastrolia:2006ki,Forde:2007mi,Ossola:2006us,Ellis:2007br,Kilgore:2007qr,Giele:2008ve,Berger:2008sj,Badger:2008cm,Ellis:2008ir}, recently reviewed in~\cite{Berger:2009zb,Britto:2010xq,Ellis:2011cr}, depend on the 
knowledge of a small, canonical set of master integrals, in which any amplitude can be expanded 
linearly with coefficients that are  rational functions of the kinematic invariants \cite{Brown:1952eu, 
Melrose:1965kb, Passarino:1978jh, Hooft1978xw, vanNeerven:1983vr, Stuart:1987tt,Stuart:1989de,vanOldenborgh:1989wn,Bern:1992em, Bern:1993kr,Fleischer:1999hq,Binoth:1999sp,Denner:2002ii,Duplancic:2003tv,Denner:2005nn,Ellis:2007qk}.  
Generalized cuts of a loop integral are defined as the singularities in the limit of sets of virtual 
momenta going on shell:  propagators are replaced by the delta functions of the on-shell conditions.  
For certain cuts, complex-valued momenta must be considered in order to solve all on-shell 
conditions.
The coefficients of the master integrals are then obtained by matching generalized cuts of the loop 
amplitude and the master integrals.

Unitarity methods work most beautifully when all internal particles are massless. 
 If massive particles 
are involved, the same cuts can be used to produce coefficients of most of the master integrals without any new conceptual difficulties. 
More significantly, there are some additional master integrals whose cuts are more difficult to solve.
 In the case of QCD processes involving  $n$ massive quarks with 
masses
$m_{i=1,\ldots,n}$,   the  additional master integrals are the tadpoles $A_0(m_i)$, and the ``on-shell bubbles'' $B_0(m_i^2;m_i,0)$ and $B_0(0;m_i,m_i)$.
 The massless counterparts of these integrals vanish in dimensional regularization.    These bubbles 
 are called on-shell because they are in the momentum channel of a single on-shell external 
 particle, either the new massive particle or any of the massless ones.  There are several proposals 
 for computing some of these  coefficients~\cite{Ossola:2006us, Mastrolia:2010nb,Ellis:2011cr,Bern:
 1995db, Badger:2008za,Britto:2009wz,Britto:2010um} 
either numerically or analytically, with or without unitarity cuts.

Here, in the context of renormalization, we will be concerned with the tadpole integral and the bubble $B_0(m^2;m,0)$, which has exactly the shape of the  external leg correction diagram of the massive particle.  
Renormalized amplitudes  also need the differentiated bubble, $B'_0(m^2;m,0)$. The derivative is taken with respect to the first argument.

External leg correction diagrams contain an internal on-shell propagator, which implies the singularity of their cuts, and hence the singularity of the cuts of the amplitude.
This problem was 
raised and nicely addressed in \cite{Ellis:2008ir,Ellis:2011cr}, in 
the context of a numerical algorithm.  The solution given in those papers was to omit the problematic 
contributions, taking care with the associated breaking of 
gauge invariance.

Here, we present a different solution.  In the spirit of the unitarity method, we want to keep the 
ingredients of the cut as complete amplitudes, without discarding any  contributions.  We must 
therefore also include the corresponding counterterms. To 
  regularize the singularity arising from the  cut of the external leg correction diagram, we 
  perform an off-shell continuation of the cut momentum.  The divergence is cancelled between the 
  cut loop diagram and the cut counterterm diagram, so we retain only the finite terms in the 
  expansion around the on-shell kinematics.
 The examples given here involve 4-dimensional cuts, although the formalism is equally valid in $D$ dimensions.
 We are also able to sum over physical polarization states only, in the spinor-helicity formalism.

The paper is organized as follows.  In Section 2, we present the theoretical analysis.  We define the 
off-shell continuation and show the cancellation of divergences for both bubble and tadpole
 coefficients from double and single cuts, respectively.   In Section 3, we illustrate the method with
  two examples computed with Feynman diagrams, namely $H \to b \bar b$ and $q \bar q \to t \bar t$.  
  In Section 4, we illustrate the method in the spinor-helicity formalism for one of the partial 
  amplitudes in $\bar t t \to gg$.  For this calculation, we make use of compact analytic formulas for 
  tree amplitudes.  In Section 5, we summarize the algorithm.  Appendix A lists the master integrals 
  and some useful reduction formulas, and Appendix B gives identities and formulas for the massive 
  spinor formalism used in Section 4.

\section{Description of the method}
\label{Sec:Wavefunction}
   
The unitarity cut of the amplitude of a  process with massive external particles should give information about the coefficients of the 
on-shell Green's function. Unfortunately the external leg correction diagrams are singular, because the propagator opposite 
the external leg carries the same momentum and is therefore also on shell.

In order to apply finite unitarity cuts, we perform 
an off-shell continuation by doing a double momentum shift.  We shift the massive external momentum $k$ and one other external momentum $r$, to preserve momentum conservation.   Assuming that both momenta are 
outgoing, the  shift may be written as 
\bea
k \to \hat k = k +   \xi \bar k, \qquad 
r \to \hat r = r -  \xi \bar k.
\label{shift}
\eea
We choose the momentum $\bar k$  such that
$r \cdot \bar k =  \bar k^2=0$ and  $2(k \cdot \bar k)  \neq0$.  
The momentum $k$  is on shell:   $k^2=m^2$, while $\hat{k}$ is off shell.  The momentum $\hat{r}$ is still on shell after the shift.
 With this shift, the propagator opposite the external leg diverges as $1/\xi$; thus the full amplitude diverges the same way. 
 The cuts are calculated in terms of tree amplitudes. For the double cut, one has simply the three-point 
 interaction $\mathcal{M}_R$, now continued off shell, and  an $(n+1)$-point tree amplitude $\mathcal{M}_L$, on shell 
 but depending on the parameter $\xi$.  In the case of the single cut we have a single $(n+2)$-point on-shell tree amplitude $\mathcal{M}_T$, which is $\xi$-dependent. We 
 expand $\mathcal{M}_L$ and $\mathcal{M}_T$  up to  first order in $\xi$. The coefficients of this expansion could be obtained  directly from  a closed form for the full amplitude.  Alternatively, separate 
 recursive constructions could be used to generate them independently.

The external leg correction diagrams are exactly cancelled 
by the corresponding counterterms, provided that the  renormalization constants are defined in the on-shell scheme.  The   
counterterm diagrams  are constructed from the renormalization constants and the various tree-level off-shell currents.  The renormalization constants are known in terms of the master integrals. The 
other ingredients needed for this step are the various tree-level off-shell currents.  An expansion in $\xi$ is performed here as well, and the divergent part is simply discarded since it is guaranteed to 
cancel the cut loop diagram.   The off-shell currents are gauge-dependent.  In the sum of all parts, gauge invariance is restored by construction:  coefficients of master integrals are gauge invariant, and 
we have only added zero in the form of the external leg correction plus its counterterm.

After the cancellation of the divergent contributions, the on-shell limit is reached by setting  $\xi =0$.  Therefore, at every stage of the procedure, 
we systematically neglect terms of $\mathcal{O}(\xi)$.

In the remainder of this  section,  we 
demonstrate the cancellation of divergences in the cut 
between loop Feynman diagrams and the counterterm. Our conventions for the master integrals and their cuts are given in Appendix \ref{App:MI}.

\subsection{Bubbles from double cut}
\label{SSSec:BubDC}
  
\FIGURE[t]{
\unitlength=0.44bp%
%
%
%
\begin{feynartspicture}(300,300)(1,1)
\FADiagram{}
\FAProp(0.,14.)(6.5,10.)(0.,){/Straight}{0}
\FAProp(0.,12.)(6.5,10.)(0.,){/Straight}{0}
\FAProp(0.,10.)(6.5,10.)(0.,){/Straight}{0}
\FAProp(0.,8.)(6.5,10.)(0.,){/Straight}{0}
\FAProp(0.,6.)(6.5,10.)(0.,){/Straight}{0}
\FAProp(21.,10.)(16.5,10.)(0.,){/Straight}{-1}
\FALabel(21.25,11.07)[b]{\scriptsize $t  (\hat k, c)$}
\FAProp(6.5,10.)(11.,10.)(0.,){/Straight}{1}
\FALabel(9.25,11.07)[b]{\scriptsize $\mathcal P$}
\FAProp(16.5,10.)(11.,10.)(-0.8,){/Straight}{-1}
\FAProp(16.5,10.)(11.,10.)(0.8,){/Cycles}{0}
\FAVert(6.5,10.){-1}
\FALabel(6.5,10.)[]{\scriptsize $\widehat{\mathcal{A}}$}
\FAVert(16.5,10.){0}
\FAVert(11.,10.){0}
\FAProp(13.75,15.)(13.75,5.)(0.,){/ScalarDash}{0}
\FALabel(11.,0.)[b]{$\Delta_{2,\hat k} \mathcal{M}$}
\end{feynartspicture}
\hspace{0.2cm} 
\begin{feynartspicture}(300,300)(1,1)
\FADiagram{}
\FAProp(0.,14.)(6.5,10.)(0.,){/Straight}{0}
\FAProp(0.,12.)(6.5,10.)(0.,){/Straight}{0}
\FAProp(0.,10.)(6.5,10.)(0.,){/Straight}{0}
\FAProp(0.,8.)(6.5,10.)(0.,){/Straight}{0}
\FAProp(0.,6.)(6.5,10.)(0.,){/Straight}{0}
\FAProp(11.,10.)(16.,14.)(0.,){/Cycles}{0}
\FALabel(19.3,13.2)[t]{\scriptsize $g (\ell, A)$}
\FAProp(11.,10.)(16., 6.)(0.,){/Straight}{1}
\FALabel(20.,6.8)[b]{\scriptsize $t (\hat k-\ell, c')$}
\FAProp(6.5,10.)(11.,10.)(0.,){/Straight}{1}
\FALabel(9.25,11.07)[b]{\scriptsize $\mathcal P$}
\FAVert(6.5,10.){-1}
\FALabel(6.5,10.)[]{\scriptsize $\widehat{\mathcal{A}}$}
\FAVert(11.,10.){0}
\FALabel(10.,0.)[b]{ $\mathcal{M}_L$}
\end{feynartspicture}
\hspace{0.2cm}  
\begin{feynartspicture}(300,300)(1,1)
\FADiagram{}
\FAProp(0.,14.)(7.5,10.)(0.,){/Cycles}{0}
\FALabel(3.37021,11.6903)[tr]{\scriptsize $g (\ell, A)$}
\FAProp(0.,6.)(7.5,10.)(0.,){/Straight}{1}
\FALabel(4.12979,6.69032)[tl]{\scriptsize $t (\hat k-\ell, c')$}
\FAProp(7.5,10.)(15.,10.)(0.,){/Straight}{1}
\FALabel(13.25,11.07)[b]{\scriptsize $t (\hat k,c)$}
\FAVert(7.5,10.){0}
\FALabel(7.,0.)[b]{ $\mathcal{M}_R$}
\end{feynartspicture}
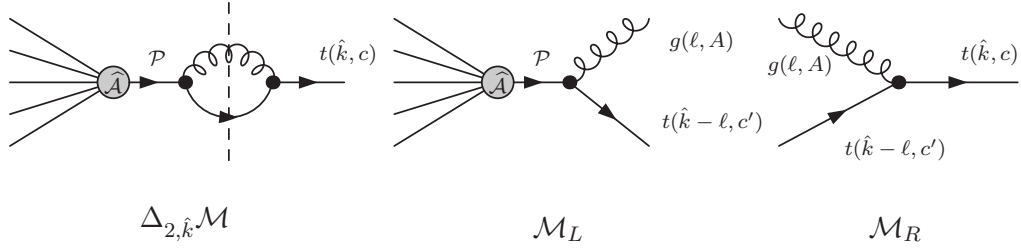
\caption{Double cut of the external leg correction diagram, and the left and right tree-level amplitudes. The cut momenta are $\ell$ and $k-\ell$.  Color information is indicated by $c$ and $c'$.  The massive propagator giving the on-shell divergence is denoted by $\mathcal{P}$.}
\label{Fig:DoubleCut}
} 
  
Consider the double cut of the external leg correction diagram for a massive fermion, 
as shown in  Figure~\ref{Fig:DoubleCut}. 
Let $k$ denote the outgoing momentum of the external fermion, and $\ell$ and $k-\ell$ the momenta of the cut gluon and cut fermion, respectively.  

The shifted massive propagator $\mathcal{P}$ is
\bea
\mathcal{P} = \frac{i(m+\slashed{k}+  \xi \slashed{\bar k})}{(k+  \xi \bar k)^2-m^2 } =   \frac{i(m+\slashed{k}+  \xi \slashed{\bar k})}{  \xi \gamma },
\eea
where $\gamma$ is defined by
\bea
\gamma \equiv 2 k \cdot \bar k.
\eea
The tree-level amplitudes $\mathcal{M}_L$ and $\mathcal{M}_R$
depicted in Figure~\ref{Fig:DoubleCut} read as follows:
\begin{eqnarray}
\mathcal{M}_L &=&  \frac{ g \, T^A_{c' c''}  }{(k+  \xi \bar k)^2-m^2 } \; \left ( \bar u_{k+   \xi \bar k-\ell} \,  \slashed{\varepsilon}^{\ast}_{\ell} \,
(m+\slashed{k}+   \xi \slashed{\bar k} ) \, \widehat{\cal A}_{c'' c_{\rm ext}}  \right ), \nonumber \\
\mathcal{M}_R &=& -i\,g\,T^A_{cc'}   \left ( \bar u_{k} \,  \slashed{\varepsilon}_{\ell} \, u_{k 
+  \xi \bar k-\ell} \right ).
\label{Mleftright}
\end{eqnarray}
Here $\widehat{\cal A}_{c'' c_{\rm ext}}$ denotes the remaining parts of the diagram, including 
the external legs on the left and all color-flow information.\footnote{
Since it depends on  $\hat k$ and $\hat r$, $\widehat{\cal A}_{c'' c_{\rm ext}}$ depends  on $  \xi$.}  
Specifically, $c''$ denotes the color of the propagator $\mathcal{P}$.

Note that the external massive spinor $u_k$ is {\em not} being shifted.   Its shift is possible but  unnecessary,  since  terms arising from the $\mathcal{O}(\xi)$ contribution of $u_k$  drop out  once the unrenormalized amplitude and the counterterm diagrams are summed together.
 We are performing the complete momentum shift on $\mathcal{M}_L$, supplemented by the 
 specific instructions given above for how to continue the propagator $\mathcal{P}$ and the vertex
 $\mathcal{M}_R$.

Multiplying these expressions, we find that the double cut including the sum over polarization states, in Feynman gauge 
where $\sum \varepsilon_\mu \varepsilon^{\ast}_\nu = -g_{\mu\nu}$, is 
\bean
\frac{2 i g^2 C_F}{ \xi   \gamma} \int d \mu_{2, \hat k} \Big [
(2 m^2 - \xi   \gamma)  \left ( \bar u_k \, \widehat{\mathcal{A}}_{cc_{\rm ext}}  \right ) +
 \left ( \bar u_k \, \slashed{\ell} \, (m+\slashed k +   \xi \slashed{\bar k})\,   \widehat{\mathcal{A}}_{cc_{\rm ext}}  \right ) +
 m   \xi \left ( \bar u_k \, \slashed{\bar k} \, \widehat{\mathcal{A}}_{cc_{\rm ext}}  \right ) 
\Big ].
\eean
The double-cut integration measure $d \mu_{2, \hat k}$ is defined in equation (\ref{2cut-lips}), and
the Dynkin index of the fundamental representation of $SU(N)$ 
is   
\begin{displaymath}
C_F =  \frac{N^2-1}{2N}.
\end{displaymath} 
It accounts for the internal color sum, leaving a Kronecker delta function allowing us to replace $c''$ by $c$ in $\widehat{\mathcal{A}}_{c''c_{\rm ext}}$.

Using integral reduction (see~(\ref{Eq:ResB0})),  we get the bubble part of the diagram:
\bean
\frac{g^2 C_F}{16 \pi^2  \xi \gamma} \Big [
4m^2 \left ( \bar u_k \, \widehat{\mathcal{A}}_{cc_{\rm ext}}  \right )  +
2  \xi m \left ( \bar u_k \, \slashed{\bar k} \, \widehat{\mathcal{A}}_{cc_{\rm ext}}  \right ) 
\Big] \, B_0(m^2+  \xi \gamma;m,0).
\label{Eq:Bub0}
\eean
Finally, we expand around $  \xi =0$, using
\begin{eqnarray}
B_0(m^2+  \xi \gamma;m,0) &=&  B_0(m^2;m,0)+ \xi  \gamma  B'_0(m^2;m,0) 
\end{eqnarray}
and  
\begin{eqnarray}
\widehat{\mathcal{A}}_{cc_{\rm ext}}  \equiv  {\mathcal{A}}_{cc_{\rm ext}} +\xi {\mathcal{A}}'_{cc_{\rm ext}}.
\end{eqnarray}
The result is that the bubble parts of the shifted diagram are given by
\begin{eqnarray}
\mathcal{M}_B &=& \frac{g^2 C_F}{16 \pi^2   \xi \gamma} \Bigg \{    
4 m^2   \xi \gamma \left (  \bar u_k \,{\mathcal{A}}_{cc_{\rm ext}}  \right )  B'_0(m^2;m,0) \nonumber 
\\
&+&   \bigg [ 
4 m^2 \left (  \bar u_k \,{\mathcal{A}}_{cc_{\rm ext}}  \right )  +
4m^2   \xi  \left (  \bar u_k \,  {\mathcal{A}}'_{cc_{\rm ext}}  \right ) +
2m   \xi  \left (  \bar u_k \,\slashed{\bar k} \, {\mathcal{A}}_{cc_{\rm ext}}  \right )
\bigg ] B_0(m^2;m,0)
  \Bigg \}  .
 \label{Eq:DCexpanded}
\end{eqnarray}

When the amplitudes are written  in the spinor-helicity formalism \cite{Berends:1981rb,De Causmaecker:1981bg,Kleiss:1985yh,Xu:1986xb,Gunion:1985vca}, the procedure described  above must be modified. Indeed, when using spinors, 
the completeness relation for polarization vectors is that of
a light-like axial gauge rather than Feynman gauge: 
\begin{equation}
\sum_{\lambda=\pm} \varepsilon_\mu \varepsilon^\ast_\nu  = -g_{\mu \nu} + \frac{\ell_\mu q_\nu + \ell_\nu q_\mu }{\ell \cdot q}.
\label{Eq:Proj1}
\end{equation}
Here  $\ell$ is the momentum of the gluon, and $q$ is an arbitrary light-like ``reference'' momentum.  The reference momentum (chosen independently for every gluon in the process) is needed to express polarization vectors in terms of spinors.  Using eq.~(\ref{Eq:Proj1}), we see that the double cut
gets an extra $\mathcal{O}(\xi^0)$ contribution of the form
\begin{eqnarray}
&& -\frac{i g^2 C_F}{\xi \gamma}   \int d \mu_{2,\hat k} 
\left[
\frac{
\left (\bar u_{k}\, \slashed{\ell} \, u_{k +   \xi \bar k -\ell} \right) 
\left( \bar u_{k +   \xi \bar k - \ell} \, \slashed{q} \, (m+ \slashed{k} 
+   \xi \slashed{\bar k} 
 ) \, \widehat{\mathcal{A}}_{c c_{\rm ext}} \right )
}{q\cdot \ell} \right.
\label{axga1} \\
&& \left. \qquad \qquad \qquad \qquad 
+   \frac{
\left (\bar u_{k}\, \slashed{q} \, u_{k +   \xi \bar k -\ell} \right) 
\left( \bar u_{k +   \xi \bar k - \ell} \, \slashed{\ell} \, (m+ \slashed{k} 
+   \xi \slashed{\bar k} 
 ) \, \widehat{\mathcal{A}}_{c c_{\rm ext} }  \right )
}{q\cdot \ell}  \right]. \nonumber
\end{eqnarray}

The second of these terms vanishes in the sum over diagrams,
\begin{equation}
\sum_{\widehat{\mathcal A}}  \;  \frac{1}{\xi \gamma}\left( \bar u_{k +   \xi \bar k - \ell} \, \slashed{\ell} \, (m+ \slashed{k} 
+   \xi \slashed{\bar k} 
 ) \, \widehat{\mathcal{A}}_{c c_{\rm ext}} \right ),
\end{equation}
 due to the Ward identity with the on-shell cut gluon.

The remaining first term in (\ref{axga1}) is equal to
\begin{eqnarray}
 - \frac{i g^2 C_F}{\gamma }    \int d \mu_{2,\hat k} \;   \frac{
\bar u_{k}\, \slashed{\bar k} \,  (m  + \slashed{k} 
+   \xi \slashed{\bar k} - \slashed{\ell} ) \, \slashed{q} \, (m+ \slashed{k} 
+   \xi \slashed{\bar k} 
 ) \,  \widehat{\mathcal{A}}_{c c_{\rm ext}} 
}{q\cdot \ell}, 
\end{eqnarray}
which gives a   contribution to the bubble part of the amplitude, (\ref{Eq:Bub0}), that is 
\begin{equation}
\frac{g^2 C_F}{16 \pi^2 \gamma} \frac{\bar u_k \,  \slashed{\bar k} \, \slashed{k} \, \slashed{q}  (m+\slashed{k})   \, {\mathcal{A}}_{c c_{\rm ext}}   }{q \cdot k}  B_0(m^2;m,0).
\label{Eq:ExtraDouble}
\end{equation}
This new contribution does not affect the divergent part of the bubble coefficient.

\FIGURE[t]{
\unitlength=0.44bp%
%
%
%
\begin{feynartspicture}(300,300)(1,1)
\FADiagram{}
\FAProp(0.,14.)(6.5,10.)(0.,){/Straight}{0}
\FAProp(0.,12.)(6.5,10.)(0.,){/Straight}{0}
\FAProp(0.,10.)(6.5,10.)(0.,){/Straight}{0}
\FAProp(0.,8.)(6.5,10.)(0.,){/Straight}{0}
\FAProp(0.,6.)(6.5,10.)(0.,){/Straight}{0}
\FAProp(21.,10.)(16.5,10.)(0.,){/Straight}{-1}
\FALabel(21.25,11.07)[b]{\scriptsize $t  (\hat k, c)$}
\FAProp(6.5,10.)(11.,10.)(0.,){/Straight}{1}
\FALabel(9.25,11.07)[b]{\scriptsize $\mathcal P$}
\FAProp(16.5,10.)(11.,10.)(-0.8,){/Straight}{-1}
\FAProp(16.5,10.)(11.,10.)(0.8,){/Cycles}{0}
\FAVert(6.5,10.){-1}
\FAVert(16.5,10.){0}
\FAVert(11.,10.){0}
\FALabel(6.5,10.1)[]{\scriptsize $\widehat{\mathcal{A}}$}
\FAProp(13.75,10.5)(13.75,5.)(0.,){/ScalarDash}{0}
\FALabel(11.,0.)[b]{$\Delta_{1,\hat k} \mathcal{M}$}
\end{feynartspicture}
\hspace{0.9cm} 
\begin{feynartspicture}(300,300)(1,1)
\FADiagram{}
\FAProp(0.,14.)(6.5,10.)(0.,){/Straight}{0}
\FAProp(0.,12.)(6.5,10.)(0.,){/Straight}{0}
\FAProp(0.,10.)(6.5,10.)(0.,){/Straight}{0}
\FAProp(0.,8.)(6.5,10.)(0.,){/Straight}{0}
\FAProp(0.,6.)(6.5,10.)(0.,){/Straight}{0}
\FAProp(21.,10.)(16.5,10.)(0.,){/Straight}{-1}
\FALabel(21.25,11.07)[b]{\scriptsize $t  (\hat k, c)$}
\FAProp(6.5,10.)(11.,10.)(0.,){/Straight}{1}
\FALabel(9.25,11.07)[b]{\scriptsize $\mathcal P$}
\FAProp(16.5,10.)(11.,10.)(0.,){/Cycles}{0}
\FAProp(16.5,10.)(21.,6.)(0.,){/Straight}{-1}
\FALabel(23.7,7.)[b]{\scriptsize $t  (\hat k-\ell, c')$}
\FAProp(11.,10.)(6.5,6.)(0.,){/Straight}{1}
\FALabel(11.4,6.)[b]{\scriptsize $t  (\hat k-\ell, c')$}
\FAVert(6.5,10.){-1}
\FAVert(16.5,10.){0}
\FAVert(11.,10.){0}
\FALabel(6.5,10.1)[]{\scriptsize $\widehat{\mathcal{A}}$}
\FALabel(10.5,0.)[b]{$\mathcal{M}_T$}
\end{feynartspicture}
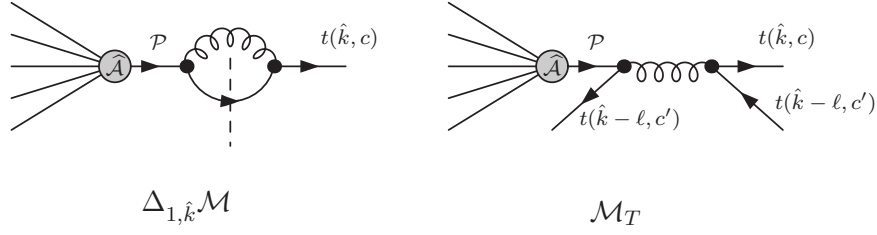
\caption{Single cut of the external leg correction diagram.}
\label{Fig:SingleCut}
}

\subsection{Tadpole from single cut}
  
Now we consider the single cut of the massive propagator of the external leg correction diagram.
See Figure~\ref{Fig:SingleCut}. \footnote{We do not compute the single cut of the gluon propagator  since the massless  tadpole vanishes in dimensional regularization. } The ingredients are nearly the same as in the double cut, except that the polarization sum for the 
gluon is replaced by the propagator $-ig_{\mu\nu}/\ell^2$. The tree-level amplitude $\mathcal{M}_{T}$ is 
\begin{eqnarray}
\mathcal{M}_{T} &=&-\frac{ g^2 C_F}{ \ell^2   \xi \gamma} 
\left (\bar u_{k}\, \gamma^\mu \, u_{k +   \xi \bar k -\ell} \right) 
\left( \bar u_{k +   \xi \bar k - \ell} \, \gamma_\mu \, (m+ \slashed{k} 
+   \xi \slashed{\bar k} 
 ) \, \widehat{\mathcal{A}}_{c'c_{\rm ext}} \right ).
\end{eqnarray}
The single cut $\Delta_{1, \hat k} \mathcal{M}$ reads as follows:
\bean
-\frac{2\, g^2 C_F}{   \xi \gamma}\int d \mu_{1,\hat k}  \left[\frac{
(2 m^2 -  \xi \gamma)  \left ( \bar u_k \, \mathcal{A}_{cc_{\rm ext}}  \right ) +
 \left ( \bar u_k \, \slashed{\ell} \, (m+\slashed k +   \xi \slashed{\bar k})\,   \mathcal{A}_{cc_{\rm ext}}  \right ) +
 m   \xi \left ( \bar u_k \, \slashed{\bar k} \, \widehat{\mathcal{A}}_{cc_{\rm ext}}  \right )}{\ell^2} 
\right ].
\eean
The single-cut integration measure $d \mu_{1, \hat k}$ is defined in equation (\ref{1cut-lips}).
The integral reduction relations~(\ref{Eq:ResA0}) allow us to compute the tadpole portion  of the external leg correction diagram.  The result is
\bea
 \frac{g^2 C_F}{16 \pi^2} \left [  \frac{
(2 m^2   \xi \gamma )  \left (  \bar u_k \,  \widehat{\mathcal{A}}_{cc_{\rm ext}}   \right)+
  \xi m  \left (  \bar u_k \, \slashed{\bar k} \,   \widehat{\mathcal{A}}_{cc_{\rm ext}}   \right) }{
   \xi\gamma (m^2+  \xi\gamma)
} \right ] \; A_0(m).
\eea
Expanding in  $  \xi \to 0$,  we get
\be
\mathcal{M}_A =  \frac{g^2 C_F}{16 \pi^2}  \bigg [
\left ( \frac{2}{  \xi\gamma} - \frac{1}{m^2} \right )   \left (  \bar u_k \, {\mathcal{A}}_{cc_{\rm ext}}   \right) + 
\frac{1}{\gamma m}   \left (  \bar u_k \,  \slashed{\bar k} \, {\mathcal{A}}_{cc_{\rm ext}}   \right) 
+\frac{2}{\gamma} \left (  \bar u_k \, {\mathcal{A}}'_{cc_{\rm ext}}   \right) 
 \bigg ]\;  A_0(m) .
 \label{Eq:SCexpanded}
\ee

\FIGURE[t]{
\unitlength=0.44bp%
\hspace{3cm}  
\begin{feynartspicture}(300,300)(1,1)
\FADiagram{}
\FAProp(0.,14.)(6.5,10.)(0.,){/Straight}{0}
\FAProp(0.,12.)(6.5,10.)(0.,){/Straight}{0}
\FAProp(0.,10.)(6.5,10.)(0.,){/Straight}{0}
\FAProp(0.,8.)(6.5,10.)(0.,){/Straight}{0}
\FAProp(0.,6.)(6.5,10.)(0.,){/Straight}{0}
\FAProp(13.,10.)(18.5, 10.)(0.,){/Straight}{1}
\FALabel(18.,11.)[7]{\scriptsize $t (\hat k, c)$}
\FAProp(6.5,10.)(13.,10.)(0.,){/Straight}{1}
\FALabel(11.25,11.07)[b]{\scriptsize $\mathcal P$}
\FAVert(6.5,10.){-1}
\FALabel(6.5,10.)[]{\scriptsize $\widehat{\mathcal{A}}$}
\FAVert(13.,10.){2}
\FALabel(10.,0.)[b]{ $\mathcal{M}^{\rm ct}$}
\end{feynartspicture}
\hspace{3cm}  

\caption{Counterterm diagram.}
\label{Fig:CTdiag}
}

\subsection{Cancellation against the counterterm}
\label{SSec:Cancellation}
Now we write the external leg counterterm, $\mathcal{M}^{\rm ct}$, depicted in Figure~\ref{Fig:CTdiag}, in terms of the master integrals, and we show that the divergent part is exactly cancelled by the divergence of the external leg correction diagram.

 The contribution of $\mathcal{M}^{\rm ct}$ is
\bea
\mathcal{M}^{\rm ct} = -\frac{1}{  \xi\gamma} \Big( 
\bar u_k \,  \left (  \slashed{k}  \delta Z_{\psi} +  \xi \slashed{\bar k}  \delta Z_{\psi}-m\delta Z_\psi - m \delta Z_m\right ) \,
 (\slashed{k}+   \xi\slashed{\bar k}+ m)\, \widehat{\cal A}_{c c_{\rm ext}} 
\Big) .
\label{Eq:CT0}
\eea
In the on-shell scheme,  the  renormalization constants $\delta Z_m$ and $\delta Z_\psi$ are expressed in terms of master integrals as
\begin{eqnarray}
\delta Z_m &=&  -{g^2 C_F \over 16\pi^2} \left[ {A_0(m) \over m^2} + 2 B_0(m^2;m,0) \right], \nonumber \\
\delta Z_\psi &=& -{g^2 C_F \over 16\pi^2} \left[{A_0(m) \over m^2} - 4 m^2 B'_0(m^2;m,0) \right].
\end{eqnarray}
Therefore, the counterterm diagram is
\be
\mathcal{M}^{\rm ct} =  \mathcal{M}^m + \mathcal{M}^{\psi},  
\label{Eq:CTexpanded0}
\ee
where $\mathcal{M}^m$ is given by
\begin{eqnarray} 
\mathcal{M}^m &=& 
-\frac{g^2 C_F}{16 \pi^2} \Bigg \{  
\bigg [ 
 \frac{2}{\xi   \gamma}A_0(m)  + \frac{4m^2}{\xi   \gamma}  B_0(m^2;m,0) \bigg  ]   \left (  \bar u_k \, \widehat{\mathcal{A}}_{cc_{\rm ext}}   \right)    
  \nonumber \pagebreak[1]  \\
&& \qquad \qquad 
+ \left [ \frac{1}{\gamma m} A_0(m) +\frac{2m}{\gamma} B_0(m^2;m,0) \right ]
\left (  \bar u_k \, \slashed{\bar k} \, \widehat{\mathcal{A}}_{cc_{\rm ext}}   \right ) \Bigg\}   \nonumber 
\pagebreak[1]  \\ 
&\simeq& -\frac{g^2 C_F}{16 \pi^2} \Bigg \{  
\bigg [ 
 \frac{2}{\xi   \gamma}A_0(m)  + \frac{4m^2}{\xi   \gamma}  B_0(m^2;m,0) \bigg  ]   \left (  \bar u_k \, {\mathcal{A}}_{cc_{\rm ext}}   \right)     \nonumber \pagebreak[1]  \\
&& \qquad \qquad 
 +
\bigg [ 
 \frac{2}{ \gamma}A_0(m)  + \frac{4m^2}{ \gamma}  B_0(m^2;m,0) \bigg  ]   \left (  \bar u_k \, {\mathcal{A}}'_{cc_{\rm ext}}   \right)     \nonumber \pagebreak[1]  \\
 && \qquad \qquad 
 +
\left [ \frac{1}{\gamma m} A_0(m) +\frac{2m}{\gamma} B_0(m^2;m,0) \right ]
\left (  \bar u_k \, \slashed{\bar k} \, {\mathcal{A}}_{cc_{\rm ext}}   \right ) 
\Bigg\},
\label{Eq:CTexpanded2}
\end{eqnarray}
while  $\mathcal{M}^{\psi}$ is given by
\begin{eqnarray}
\mathcal{M}^{\psi} &=&
-\frac{g^2 C_F}{16 \pi^2} \Bigg \{  
\bigg [ 
  -\frac{1}{m^2}A_0(m) +   4m^2 B'_0(m^2;m,0)  \bigg  ]   \left (  \bar u_k \, \widehat{\mathcal{A}}_{cc_{\rm ext}}   \right)    
\Bigg\} \nonumber  \pagebreak[1]  \\
&\simeq& 
-\frac{g^2 C_F}{16 \pi^2} \Bigg \{  
\bigg [ 
  -\frac{1}{m^2}A_0(m) +   4m^2 B'_0(m^2;m,0)  \bigg  ]   \left (  \bar u_k \, {\mathcal{A}}_{cc_{\rm ext}}   \right)   
\Bigg\}. 
\label{Eq:CTexpanded1}
\end{eqnarray}
When the spinor-helicity  formalism  is used, the extra contribution~(\ref{Eq:ExtraDouble}) is 
accounted for by  adding the following term to eq.~(\ref{Eq:CT0}):
\begin{eqnarray}
\label{Eq:CTexpanded3}
\mathcal{M}^{k} &=& -\frac{1}{\xi \gamma}  \bar u_k  \bigg  [  (\slashed{k}+ \xi \slashed{\bar k} - m ) \, 
(\slashed{k} + \xi \slashed{\bar k} ) \slashed{q} \, \delta Z_{k}' \bigg ] (\slashed{k} + \xi \slashed{\bar 
k} + m)  \widehat{\mathcal{A}}_{c c_{\rm ext}},
\end{eqnarray}
where
\begin{eqnarray}
\delta Z_{k}'  &=& \frac{g^2 C_F}{16 \pi^2}  \frac{ B_0(m^2;m,0)}{q\cdot k}.
\end{eqnarray}

Having expanded the counterterm around $  \xi =0$ at $\mathcal{O}(  \xi^0)$,
it is straightforward to check the cancellation of the terms proportional to  $  \xi^{-1}$  and the ones 
which depend on $\bar k$ once  $\mathcal{M}_A$,  $\mathcal{M}_B $, and 
$\mathcal{M}^{\rm ct}$ are added together.  The  sum is identically zero, since in the on-shell scheme the external leg correction diagram 
is exactly compensated  by  the external leg counterterm.  The actual contribution to the tadpole and bubble coefficients comes from the other diagrams in the full amplitude. They are finite in 
$\xi$, thus no $\bar k$ dependence arises in the $\xi \to 0$ limit.

\section{Examples}
\label{Sec:Examples}
  
\FIGURE[t]{
\unitlength=0.42bp%
%
%
%
%
\begin{feynartspicture}(300,300)(1,1)
\FADiagram{}
\FAProp(0.,10.)(6.5,10.)(0.,){/ScalarDash}{0}
\FALabel(3.25,9.18)[t]{\scriptsize $H(k_1)$}
\FAProp(20.,15.)(13.,14.)(0.,){/Straight}{-1}
\FALabel(16.2808,15.5544)[b]{\scriptsize $b(k_2,c_2)$}
\FAProp(20.,5.)(13.,6.)(0.,){/Straight}{1}
\FALabel(16.2808,4.44558)[t]{\scriptsize $b(k_3,c_3)$}
\FAProp(6.5,10.)(13.,14.)(0.,){/Straight}{1}
\FALabel(9.20801,13.1807)[br]{\scriptsize $b$}
\FAProp(6.5,10.)(13.,6.)(0.,){/Straight}{-1}
\FALabel(9.20801,6.81927)[tr]{\scriptsize $b$}
\FAProp(13.,14.)(13.,6.)(0.,){/Cycles}{0}
\FALabel(14.974,10.)[l]{\scriptsize $g$}
\FAVert(6.5,10.){0}
\FAVert(13.,14.){0}
\FAVert(13.,6.){0}
\end{feynartspicture}
\begin{feynartspicture}(300,300)(1,1)
\FADiagram{}
\FAProp(0.,10.)(11.,10.)(0.,){/ScalarDash}{0}
\FALabel(5.5,9.18)[t]{\scriptsize $H(k_1)$}
\FAProp(20.,15.)(11.,10.)(0.,){/Straight}{-1}
\FALabel(15.2273,13.3749)[br]{\scriptsize $b(k_2,c_2)$}
\FAProp(20.,5.)(17.3,6.5)(0.,){/Straight}{1}
\FALabel(18.9227,6.62494)[bl]{\scriptsize $b(k_3,c_3)$}
\FAProp(11.,10.)(13.7,8.5)(0.,){/Straight}{-1}
\FALabel(12.0773,8.37506)[tr]{\scriptsize $b$}
\FAProp(17.3,6.5)(13.7,8.5)(-0.8,){/Straight}{1}
\FALabel(14.4273,5.18506)[tr]{\scriptsize $b$}
\FAProp(17.3,6.5)(13.7,8.5)(0.8,){/Cycles}{0}
\FALabel(16.5727,9.81494)[bl]{\scriptsize $g$}
\FAVert(11.,10.){0}
\FAVert(17.3,6.5){0}
\FAVert(13.7,8.5){0}
\end{feynartspicture}
\begin{feynartspicture}(300,300)(1,1)
\FADiagram{}
\FAProp(0.,10.)(11.,10.)(0.,){/ScalarDash}{0}
\FALabel(5.5,9.18)[t]{\scriptsize $H(k_1)$}
\FAProp(20.,15.)(17.3,13.5)(0.,){/Straight}{-1}
\FALabel(18.3773,15.1249)[br]{\scriptsize $b(k_2,c_2)$}
\FAProp(20.,5.)(11.,10.)(0.,){/Straight}{1}
\FALabel(15.2273,6.62506)[tr]{\scriptsize $b(k_3,c_3)$}
\FAProp(11.,10.)(13.7,11.5)(0.,){/Straight}{1}
\FALabel(12.0773,11.6249)[br]{\scriptsize $b$}
\FAProp(17.3,13.5)(13.7,11.5)(-0.8,){/Straight}{-1}
\FALabel(16.5727,10.1851)[tl]{\scriptsize $b$}
\FAProp(17.3,13.5)(13.7,11.5)(0.8,){/Cycles}{0}
\FALabel(13.4273,13.8149)[br]{\scriptsize $g$}
\FAVert(11.,10.){0}
\FAVert(17.3,13.5){0}
\FAVert(13.7,11.5){0}
\end{feynartspicture}

\caption{
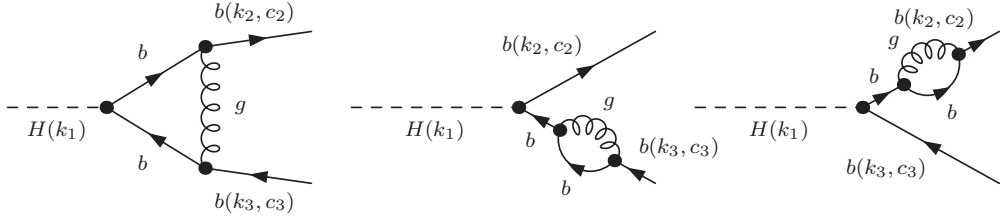
Virtual corrections to the process~(\ref{Eq:HBBpro}).  }
\label{Fig:HBBvirtual}
}

\subsection{The process $H\to b \bar b$}
\label{SSec:HBB}
As a first application, we focus on a simple process, the Higgs decay into a 
bottom--anti-bottom  pair:
\bea
H(k_1) \, \to \,  b(k_2, c_2) \, \bar b(k_3, c_3).
\label{Eq:HBBpro}
\eea
We apply the procedure described in Section~\ref{Sec:Wavefunction} to compute the
coefficients of the on-shell bubble $B_0(m_b^2;m_b,0) $, of $B'_0(m_b^2;m_b,0)$ and of the tadpole
$A_0(m_b)$ entering the NLO QCD  amplitude  
$\mathcal{M}_{H\to b \bar b}$~\cite{Braaten:1980yq,Sakai:1980fa,Inami:1980qp,Drees:1990dq}.  
We regularize
the tree-level amplitudes using the following shift:
\begin{equation}
k_2 \to \hat k_2 = k_2 +   \xi \bar k, \qquad k_3 \to \hat k_3 = k_3  -   \xi \bar k.
\end{equation}
The massless momentum $\bar k$ is such that
\begin{equation}
 \gamma_2 \equiv 2 (k_2\cdot \bar k) \neq 0, \qquad
 \gamma_3 \equiv 2 (k_3\cdot  \bar k) \neq 0,
 \end{equation}
 while $k_2$ and $k_3$ are on-shell:  $k_2^2=k_3^2=m_b^2$. Here we use a slightly 
 different shift since both the shifted momenta are off shell. In this example, the modification
 of the method  is irrelevant and streamlines the computation.\\

The coefficient $b_{B_0(m_b^2 + \gamma_2   \xi;m_b,0)}$ of  $B_0(m_b^2+\gamma_2   \xi ;m_b,0) $ 
can be  computed by means of the double cut,
\bea
\Delta_{2,\hat k_2} \mathcal{M}_{H\to b \bar b} =  \int d \mu_{2,\hat k_2} \;  \mathcal{M}_{H\to bg \bar b}(k_1, \hat  k_2-\ell,\ell, \hat k_3) \;
\mathcal{M}_{gb\to b}(\ell,\hat  k_2-\ell, \hat  k_2),
\eea
where the tree-level amplitudes on the r.h.s. are related to the processes
\begin{displaymath}
H(k_1) \, \to \,  b(k_2, c_2) \,  g(k_4, A_4) \, \bar b(k_3, c_3), \qquad 
g(k_1, A_1) \,  b(k_2, c_2)  \, \to  \,  b(k_3, c_3), 
\end{displaymath}
respectively. 
Up to $\mathcal{O}(  \xi)$, the relevant part of the double cut is given by
\begin{eqnarray}
&-& \frac{8 g^2 e m_b  \delta^{c_2}_{c_3}}{3 M_W s_{\rm w} }  \int d \mu_{2,k_2} \;   \Bigg \{
 \frac{
 m  \left ( \bar u_{k_2} \, \slashed{\ell} \, v_{k_3}  \right )  + (k_1\cdot \ell) \left ( \bar u_{k_2} \, v_{k_3}  \right )
}{(k_3+\ell)^2-m_b^2}   \nonumber \\
&+&
\frac{
 \left ( \bar u_{k_2} \, \slashed{\ell} \, ( \slashed{k_2} +\xi \slashed{\bar k}) \, v_{k_3}  \right ) + 
 m  \left ( \bar u_{k_2} ( \slashed{\ell} + \slashed{k_2} + \xi \slashed{\bar k} ) v_{k_3}  \right )-
(\gamma_2 \xi -  m_b^2)  \left ( \bar u_{k_2} \, v_{k_3}  \right )
 }{2\gamma_2 \xi}   + \cdots
\Bigg\}.
\end{eqnarray}
Using the results~(\ref{Eq:ResB0}), we get the coefficient
\begin{eqnarray} 
b_{B_0(m_b^2 +\gamma_2   \xi ;m_b,0)} &=& \frac{g^2em_b \delta^{c_2}_{c_3}i}{24 \pi^2 M_W s_{\rm w}}  \Bigg
\{ \left [  \frac{2(2m_b^2 - M_H^2)}{4m_b^2 - M_H^2}  
-\frac{\gamma_2 \xi -2 m_b^2}{\gamma_2 \xi}
\right]
\left ( \bar u_{k_2} \, v_{k_3}  \right ) 
\nonumber \\ &+&
\label{Eq:HBBb1}
  \frac{m_b (2 m_b^2 +3 \gamma_2 \xi)}{(m_b^2+  \gamma_2 \xi)\gamma_2 \xi}
  \left [ m_b  \left ( \bar u_{k_2}\,v_{k_3}  \right ) + \xi   \left ( \bar u_{k_2}\,\slashed{\bar k} \, v_{k_3}  \right ) 
 \right ] 
\Bigg \}.   
\end{eqnarray}
The parameter $e$ is the charge of the positron, while $M_W$ and  $s_{\rm w}$ are the mass of the $W$-boson
and the sine of the  weak mixing angle respectively.  \\

The coefficient $b_{B_0(m_b^2 -   \xi \gamma_3 ;m_b,0)}$ of $B_0(m_b^2-  \xi \gamma_3;m_b,0)$  can be obtained analogously 
starting from the double cut 
\bea
\Delta_{2,\hat k_3} \mathcal{M}_{H\to b \bar b} =  \int d \mu_{2,\hat k_3} \;  \mathcal{M}_{H\to bg \bar b}(k_1, \hat k_2,\ell, \hat k_3-\ell) \;
\mathcal{M}_{g\bar b\to \bar b}(\ell, \hat k_3-\ell, \hat k_3),
\eea
and using the relations~(\ref{Eq:ResB0}).  
$\mathcal{M}_{g\bar b \to \bar b}$ is the tree-level amplitude of the process
\begin{displaymath}
g(k_1, A_1) \, \bar  b(k_2, c_2)  \, \to  \,  \bar b(k_3, c_3). 
\end{displaymath}
The result is
\begin{eqnarray}
b_{B_0(m_b^2 -\gamma_3   \xi ;m_b,0)} &=& \frac{g^2em_b \delta^{c_2}_{c_3}i}{24 \pi^2 M_W s_{\rm w}} \Bigg
\{ \left [  \frac{2(2m_b^2 - M_H^2)}{4m_b^2 - M_H^2}  
- \frac{2 m_b^2+\gamma_3 \xi}{\gamma_3 \xi}
\right]
\left ( \bar u_{k_2} \, v_{k_3}  \right ) 
\nonumber \\
\label{Eq:HBBb2}
&+&
  \frac{m_b (3\gamma_3 \xi - 2 m_b^2)}{\gamma_3 \xi (m_b^2 -\gamma_3 \xi)}
  \left [\xi 
  \left ( \bar u_{k_2}\, \slashed{\bar k} \, v_{k_3}  \right )
  + m_b \left ( \bar u_{k_2}\, v_{k_3}  \right )
  \right ] \Bigg \} .
\end{eqnarray}
The coefficients of $B_0(m_b^2;m_b,0)$ and of $B'_0(m_b^2;m_b,0)$ can be obtained by expanding
\begin{equation}
b_{B_0(m_b^2 +\gamma_2   \xi ;m_b,0)} B_0(m_b^2 + \gamma_2   \xi ;m_b,0) +
b_{B_0(m_b^2 -\gamma_3   \xi ;m_b,0)} B_0(m_b^2 -\gamma_3   \xi ;m_b,0)
\end{equation}
around $\xi =0$ and  neglecting terms of   $\mathcal{O}(  \xi)$. \\

The tadpole coefficient $a_{A_0(m_b)}$ can be computed 
from the sum of two single cut diagrams,
\bea
\Delta_{1,\hat k_2} \mathcal{M}_{H\to b \bar b}  + \Delta_{1,-\hat k_3} \mathcal{M}_{H\to b \bar b}  =
 \int d \mu_{1,0} \;  \mathcal{M}_{H b \to b b \bar b}(k_1, \ell, \hat k_2, \ell, \hat k_3),
 \label{Eq:HBBsinglestart}
\eea 
provided that the loop momentum $\ell$ is chosen appropriately.\footnote{In particular in  each diagram 
of Figure~\ref{Fig:HBBvirtual} the loop momentum  has to be fixed such that the only internal propagator appearing are 
$\ell^2$, $(\ell+\hat k_2)^2 -m_b^2$, and $(\ell-\hat k_3)^2 -m_b^2$.} The tree-level amplitude on the r.h.s. is related to the process
\be
H(k_1) b(k_2, c) \, \to \,  b(k_3, c_3) \,  b(k_4, c) \, \bar b(k_5, c_5).
\label{Eq:Tree4Single}
\ee
Divergent diagrams with zero-momentum internal gluons do not contribute, owing to the color flow of the process~(\ref{Eq:Tree4Single}).
At $\mathcal{O}(  \xi)$, the part of the single cut~(\ref{Eq:HBBsinglestart}) which is relevant for the computation 
of the tadpole coefficient reads as follows:
\begin{eqnarray}
&-& \frac{4 g^2 e m_b  \delta^{c_2}_{c_3} i }{3 M_W s_{\rm w} }  \int d \mu_{1,0} \;   \Bigg \{  
\frac{
 \left ( \bar u_{k_2} \, \slashed{\ell} \, (\slashed{k_2}+ \xi \slashed{\bar k} )\, v_{k_3}  \right ) +m_b \left (  \bar u_{k_2} \, \slashed{\ell} \,  v_{k_3}  \right )
}{(\ell + k_2+\xi \bar k)^2 \, \gamma_2 \xi} \nonumber \\
&& \qquad
-\frac{
m_b  \left ( \bar u_{k_2} \, \slashed{\ell} \, v_{k_3}  \right )  -
\left ( \bar u_{k_2} \, (\slashed{k_3} - \xi \slashed{\bar k}   )\, \slashed{\ell} \,  \, v_{k_3}  \right ) 
 }{(\ell + k_3- \xi \bar k)^2 \, \gamma_3 \xi} + \cdots
\Bigg\} .
\end{eqnarray}
The coefficient $a_{A_0(m_b)} $ is obtained  using the relations~(\ref{Eq:ResA0}) and expanding 
the following  expression in $\xi$:
\begin{eqnarray}
  &~& \frac{g^2em_b \delta^{c_2}_{c_3}i}{24 \pi^2 M_W s_{\rm w}} \Bigg \{ 
  \left [ m_b \xi   \left ( \bar u_{k_2} \, \slashed{\bar k} \, v_{k_3}  \right ) +m_b^2 \left ( \bar u_{k_2} \,  v_{k_3}  \right )  \right ]
\bigg [
\frac{1}{(m^2_b + \gamma_2 \xi)\gamma_2\xi}  
 \nonumber \\
 \label{Eq:HBBa}
&-& \frac{1}{(m^2_b - \gamma_3 \xi)\gamma_3\xi} 
 \bigg ] + 
\left (  \frac{1}{\gamma_2\xi}  -     \frac{1}{\gamma_3\xi}   \right ) \left ( \bar u_{k_2} \, v_{k_3}  \right ) 
\Bigg \}.  
\end{eqnarray}

The results~(\ref{Eq:HBBb1}),~(\ref{Eq:HBBb2}),  and~(\ref{Eq:HBBa}),   have been checked  performing Passarino-Veltman decomposition of the one loop amplitude obtained from the Feynman diagrams in Figure~\ref{Fig:HBBvirtual}. This procedure 
has been performed  with the help of \verb+FeynArts+~\cite{Hahn:2000kx}, 
\verb+FormCalc+~\cite{Hahn:1998yk,Hahn:2006qw} and 
\verb+FeynCalc+~\cite{Mertig:1990an}.

We have  verified the cancellation of the $\mathcal{O}(  \xi^{-1})$ terms 
against the contribution of the two counterterm diagrams depicted in Figure~\ref{Fig:HBBct},
along the lines of  Section~\ref{SSec:Cancellation}.    In the  $\xi \to 0$ limit, the terms
containing $\bar k$ drop out.  The bubble and the tadpole coefficients are   thus independent 
 of  this unphysical momentum.

%
%
%
%
%

\FIGURE[t]{
\unitlength=0.42bp%
\scriptsize 
\begin{feynartspicture}(300,300)(1,1)
\FADiagram{}
\FAProp(0.,10.)(11.,10.)(0.,){/ScalarDash}{0}
\FALabel(5.5,9.18)[t]{$H(k_1)$}
\FAProp(20.,15.)(11.,10.)(0.,){/Straight}{-1}
\FALabel(15.2273,13.3749)[br]{$b(k_2,c_2)$}
\FAProp(20.,5.)(15.5,7.5)(0.,){/Straight}{1}
\FALabel(17.4773,5.37506)[tr]{$b(k_3,c_3)$}
\FAProp(15.5,7.5)(11.,10.)(0.,){/Straight}{1}
\FALabel(12.9773,7.87506)[tr]{$b$}
\FAVert(11.,10.){0}
\FAVert(15.5,7.5){2}
\end{feynartspicture}
\hspace{0.3cm}
\begin{feynartspicture}(300,300)(1,1)
\FADiagram{}
\FAProp(0.,10.)(11.,10.)(0.,){/ScalarDash}{0}
\FALabel(5.5,9.18)[t]{$H(k_1)$}
\FAProp(20.,15.)(15.5,12.5)(0.,){/Straight}{-1}
\FALabel(17.4773,14.6249)[br]{$b(k_2,c_2)$}
\FAProp(20.,5.)(11.,10.)(0.,){/Straight}{1}
\FALabel(15.2273,6.62506)[tr]{$b(k_3,c_3)$}
\FAProp(15.5,12.5)(11.,10.)(0.,){/Straight}{-1}
\FALabel(12.9773,12.1249)[br]{$b$}
\FAVert(11.,10.){0}
\FAVert(15.5,12.5){2}
\end{feynartspicture}

\caption{
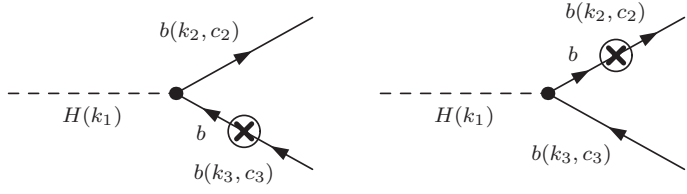
External legs  counterterm diagrams of  the process~(\ref{Eq:HBBpro}).  }
\label{Fig:HBBct}
}

\subsection{The process $q \bar q \to t \bar t$}
A less trivial example involves the process of top--anti-top production via quark--anti-quark annihilation:
\bea
q(k_1,c1)\, \bar q(k_2, c_2) \, \to \,  t(k_3, c_3)\, \bar t(k_4, c_4), \qquad (q \neq t).
\label{Eq:QQTTpro} 
\eea
This is one among the channels entering the hadronic production of a top--anti-top pair
whose NLO QCD contributions have been computed~\cite{Nason:1987xz,Nason:1989zy,Beenakker:1988bq,Beenakker:1990maa} and implemented in  \verb+MCFM+~\cite{Campbell:2000bg} and in \verb+MC@NLO+~\cite{Frixione:2003ei}.
 We have computed the coefficient of  $B_0(m_t^2;m_t,0)$, $B'_0(m_t^2;m_t,0)$ and of the tadpole $A_0(m_t)$.
 The tree-level amplitudes have been regularized using the shift
 \begin{equation}
 k_3 \to \hat k_3 = k_3 + \xi \bar k, \qquad k_4 \to \hat k_4 = k_4 - \xi \bar k.
 \end{equation}
We have used the  double cuts
\begin{eqnarray}
\label{EqQQTTcut1}
\Delta_{2,\hat k_3} \mathcal{M}_{q \bar q\to t \bar t} &=&  \int d \mu_{2,\hat k_3} \;  \mathcal{M}_{q \bar q \to t g \bar t}(k_1, k_2, \hat k_3-\ell,\ell, \hat k_4) \; \mathcal{M}_{gt\to t}(\ell, \hat k_3-\ell, \hat k_3),\\
\Delta_{2,\hat k_4} \mathcal{M}_{q \bar q\to t \bar t} &=&  \int d \mu_{2,\hat k_4} \;  \mathcal{M}_{q \bar q \to t g \bar t}(k_1, k_2, \hat k_3,\ell, \hat k_4-\ell) \; \mathcal{M}_{g \bar t\to \bar t}(\ell, \hat k_4-\ell, \hat k_4), 
\label{EqQQTTcut2}
\end{eqnarray}
the single cut
\bea
\Delta_{1, \hat k_3} \mathcal{M}_{q \bar q\to t \bar t}  + \Delta_{1,- \hat k_4} \mathcal{M}_{q \bar q\to t \bar t}  
= \int d \mu_{1,0} \;  \mathcal{M}_{q \bar q  t \to t t   \bar t}(k_1, k_2,  \ell,  \hat k_3,\ell, \hat k_4),
 \label{EqQQTTcut3}
\eea 
and the relations~(\ref{Eq:ResB0})  and~(\ref{Eq:ResA0}).
The coefficients  can be obtained performing an expansion around $\xi=0$
at $\mathcal{O}(\xi^0)$, along the lines of Section~\ref{SSec:HBB}.  The actual expression of the coefficients is  lengthy and  is omitted.
We have checked our result against  the Passarino-Veltman based computation  of the Feynman 
diagrams depicted in Figure~\ref{Fig:QQTTloop}. In the renormalized amplitude, we have checked the cancellation of the $\mathcal{O}(\xi^{-1})$
 contributions and of the terms containing the massless momentum $\bar k$.
  
\FIGURE[t]{
\unitlength=0.25bp
\scriptsize
%
%
\begin{feynartspicture}(300,300)(1,1)
\FADiagram{}
\FAProp(0.,15.)(4.,10.)(0.,){/Straight}{1}
\FALabel(1.26965,12.0117)[tr]{$q$}
\FAProp(0.,5.)(4.,10.)(0.,){/Straight}{-1}
\FALabel(2.73035,7.01172)[tl]{$q$}
\FAProp(20.,15.)(16.,13.5)(0.,){/Straight}{-1}
\FALabel(17.4558,15.2213)[b]{$t$}
\FAProp(20.,5.)(16.,6.5)(0.,){/Straight}{1}
\FALabel(17.4558,4.77869)[t]{$t$}
\FAProp(4.,10.)(10.,10.)(0.,){/Cycles}{0}
\FALabel(7.,11.77)[b]{$g$}
\FAProp(16.,13.5)(16.,6.5)(0.,){/Cycles}{0}
\FALabel(17.77,10.)[l]{$g$}
\FAProp(16.,13.5)(10.,10.)(0.,){/Straight}{-1}
\FALabel(12.699,12.6089)[br]{$t$}
\FAProp(16.,6.5)(10.,10.)(0.,){/Straight}{1}
\FALabel(12.699,7.39114)[tr]{$t$}
\FAVert(4.,10.){0}
\FAVert(16.,13.5){0}
\FAVert(16.,6.5){0}
\FAVert(10.,10.){0}
\end{feynartspicture}
%
%
\begin{feynartspicture}(300,300)(1,1)
\FADiagram{}
\FAProp(0.,15.)(4.,10.)(0.,){/Straight}{1}
\FALabel(1.26965,12.0117)[tr]{$q$}
\FAProp(0.,5.)(4.,10.)(0.,){/Straight}{-1}
\FALabel(2.73035,7.01172)[tl]{$q$}
\FAProp(20.,15.)(16.,13.5)(0.,){/Straight}{-1}
\FALabel(17.4558,15.2213)[b]{$t$}
\FAProp(20.,5.)(16.,6.5)(0.,){/Straight}{1}
\FALabel(17.4558,4.77869)[t]{$t$}
\FAProp(4.,10.)(10.,10.)(0.,){/Cycles}{0}
\FALabel(7.,11.77)[b]{$g$}
\FAProp(16.,13.5)(16.,6.5)(0.,){/Straight}{-1}
\FALabel(17.07,10.)[l]{$t$}
\FAProp(16.,13.5)(10.,10.)(0.,){/Cycles}{0}
\FALabel(12.699,12.6089)[br]{$g$}
\FAProp(16.,6.5)(10.,10.)(0.,){/Cycles}{0}
\FALabel(12.3463,6.7865)[tr]{$g$}
\FAVert(4.,10.){0}
\FAVert(16.,13.5){0}
\FAVert(16.,6.5){0}
\FAVert(10.,10.){0}
\end{feynartspicture}
%
%
\begin{feynartspicture}(300,300)(1,1)
\FADiagram{}
\FAProp(0.,15.)(4.,13.5)(0.,){/Straight}{1}
\FALabel(2.54424,15.2213)[b]{$q$}
\FAProp(0.,5.)(4.,6.5)(0.,){/Straight}{-1}
\FALabel(2.54424,4.77869)[t]{$q$}
\FAProp(20.,15.)(16.,10.)(0.,){/Straight}{-1}
\FALabel(17.2697,12.9883)[br]{$t$}
\FAProp(20.,5.)(16.,10.)(0.,){/Straight}{1}
\FALabel(18.7303,7.98828)[bl]{$t$}
\FAProp(16.,10.)(10.,10.)(0.,){/Cycles}{0}
\FALabel(13.,8.23)[t]{$g$}
\FAProp(4.,13.5)(4.,6.5)(0.,){/Cycles}{0}
\FALabel(2.93,10.)[r]{$g$}
\FAProp(4.,13.5)(10.,10.)(0.,){/Straight}{1}
\FALabel(7.301,12.6089)[bl]{$q$}
\FAProp(4.,6.5)(10.,10.)(0.,){/Straight}{-1}
\FALabel(7.301,7.39114)[tl]{$q$}
\FAVert(4.,13.5){0}
\FAVert(4.,6.5){0}
\FAVert(16.,10.){0}
\FAVert(10.,10.){0}
\end{feynartspicture}
%
%
\begin{feynartspicture}(300,300)(1,1)
\FADiagram{}
\FAProp(0.,15.)(4.,13.5)(0.,){/Straight}{1}
\FALabel(2.54424,15.2213)[b]{$q$}
\FAProp(0.,5.)(4.,6.5)(0.,){/Straight}{-1}
\FALabel(2.54424,4.77869)[t]{$q$}
\FAProp(20.,15.)(16.,10.)(0.,){/Straight}{-1}
\FALabel(17.2697,12.9883)[br]{$t$}
\FAProp(20.,5.)(16.,10.)(0.,){/Straight}{1}
\FALabel(18.7303,7.98828)[bl]{$t$}
\FAProp(16.,10.)(10.,10.)(0.,){/Cycles}{0}
\FALabel(13.,8.23)[t]{$g$}
\FAProp(4.,13.5)(4.,6.5)(0.,){/Straight}{1}
\FALabel(2.93,10.)[r]{$q$}
\FAProp(4.,13.5)(10.,10.)(0.,){/Cycles}{0}
\FALabel(7.65371,13.2135)[bl]{$g$}
\FAProp(4.,6.5)(10.,10.)(0.,){/Cycles}{0}
\FALabel(7.301,7.39114)[tl]{$g$}
\FAVert(4.,13.5){0}
\FAVert(4.,6.5){0}
\FAVert(16.,10.){0}
\FAVert(10.,10.){0}
\end{feynartspicture}
\begin{feynartspicture}(300,300)(1,1)
%
%
\FADiagram{}
\FAProp(0.,15.)(6.5,13.5)(0.,){/Straight}{1}
\FALabel(3.59853,15.2803)[b]{$q$}
\FAProp(0.,5.)(6.5,6.5)(0.,){/Straight}{-1}
\FALabel(3.59853,4.71969)[t]{$q$}
\FAProp(20.,15.)(13.5,13.5)(0.,){/Straight}{-1}
\FALabel(16.4015,15.2803)[b]{$t$}
\FAProp(20.,5.)(13.5,6.5)(0.,){/Straight}{1}
\FALabel(16.4015,4.71969)[t]{$t$}
\FAProp(6.5,13.5)(6.5,6.5)(0.,){/Straight}{1}
\FALabel(5.43,10.)[r]{$q$}
\FAProp(6.5,13.5)(13.5,13.5)(0.,){/Cycles}{0}
\FALabel(10.,15.27)[b]{$g$}
\FAProp(6.5,6.5)(13.5,6.5)(0.,){/Cycles}{0}
\FALabel(10.,5.43)[t]{$g$}
\FAProp(13.5,13.5)(13.5,6.5)(0.,){/Straight}{-1}
\FALabel(14.57,10.)[l]{$t$}
\FAVert(6.5,13.5){0}
\FAVert(6.5,6.5){0}
\FAVert(13.5,13.5){0}
\FAVert(13.5,6.5){0}
\end{feynartspicture}
%
%
\begin{feynartspicture}(300,300)(1,1)
\FADiagram{}
\FAProp(0.,15.)(6.5,13.5)(0.,){/Straight}{1}
\FALabel(3.59853,15.2803)[b]{$q$}
\FAProp(0.,5.)(6.5,6.5)(0.,){/Straight}{-1}
\FALabel(3.59853,4.71969)[t]{$q$}
\FAProp(20.,15.)(13.5,6.5)(0.,){/Straight}{-1}
\FALabel(17.9814,13.8219)[br]{$t$}
\FAProp(20.,5.)(13.5,13.5)(0.,){/Straight}{1}
\FALabel(17.8314,6.42814)[tr]{$t$}
\FAProp(6.5,13.5)(6.5,6.5)(0.,){/Straight}{1}
\FALabel(5.43,10.)[r]{$q$}
\FAProp(6.5,13.5)(13.5,13.5)(0.,){/Cycles}{0}
\FALabel(10.,15.27)[b]{$g$}
\FAProp(6.5,6.5)(13.5,6.5)(0.,){/Cycles}{0}
\FALabel(10.,5.43)[t]{$g$}
\FAProp(13.5,6.5)(13.5,13.5)(0.,){/Straight}{-1}
\FALabel(12.43,10.)[r]{$t$}
\FAVert(6.5,13.5){0}
\FAVert(6.5,6.5){0}
\FAVert(13.5,6.5){0}
\FAVert(13.5,13.5){0}
\end{feynartspicture}
%
%
\begin{feynartspicture}(300,300)(1,1)
\FADiagram{}
\FAProp(0.,15.)(4.,10.)(0.,){/Straight}{1}
\FALabel(1.26965,12.0117)[tr]{$q$}
\FAProp(0.,5.)(4.,10.)(0.,){/Straight}{-1}
\FALabel(2.73035,7.01172)[tl]{$q$}
\FAProp(20.,15.)(16.,10.)(0.,){/Straight}{-1}
\FALabel(17.2697,12.9883)[br]{$t$}
\FAProp(20.,5.)(16.,10.)(0.,){/Straight}{1}
\FALabel(18.7303,7.98828)[bl]{$t$}
\FAProp(4.,10.)(10.,10.)(0.,){/Cycles}{0}
\FALabel(7.,8.93)[t]{$g$}
\FAProp(16.,10.)(10.,10.)(0.,){/Cycles}{0}
\FALabel(13.,8.23)[t]{$g$}
\FAProp(10.,10.)(10.,10.)(10.,15.){/Cycles}{0}
\FALabel(10.,16.07)[b]{$g$}
\FAVert(4.,10.){0}
\FAVert(16.,10.){0}
\FAVert(10.,10.){0}
\end{feynartspicture}
%
%
\begin{feynartspicture}(300,300)(1,1)
\FADiagram{}
\FAProp(0.,15.)(3.,10.)(0.,){/Straight}{1}
\FALabel(0.650886,12.1825)[tr]{$q$}
\FAProp(0.,5.)(3.,10.)(0.,){/Straight}{-1}
\FALabel(2.34911,7.18253)[tl]{$q$}
\FAProp(20.,15.)(17.,10.)(0.,){/Straight}{-1}
\FALabel(17.6509,12.8175)[br]{$t$}
\FAProp(20.,5.)(17.,10.)(0.,){/Straight}{1}
\FALabel(19.3491,7.81747)[bl]{$t$}
\FAProp(3.,10.)(7.,10.)(0.,){/Cycles}{0}
\FALabel(5.,11.77)[b]{$g$}
\FAProp(17.,10.)(13.,10.)(0.,){/Cycles}{0}
\FALabel(15.,8.23)[t]{$g$}
\FAProp(7.,10.)(13.,10.)(0.8,){/Straight}{-1}
\FALabel(10.,6.53)[t]{$q'$}
\FAProp(7.,10.)(13.,10.)(-0.8,){/Straight}{1}
\FALabel(10.,13.47)[b]{$q'$}
\FAVert(3.,10.){0}
\FAVert(17.,10.){0}
\FAVert(7.,10.){0}
\FAVert(13.,10.){0}
\end{feynartspicture}
%
%
\begin{feynartspicture}(300,300)(1,1)
\FADiagram{}
\FAProp(0.,15.)(3.,10.)(0.,){/Straight}{1}
\FALabel(0.650886,12.1825)[tr]{$q$}
\FAProp(0.,5.)(3.,10.)(0.,){/Straight}{-1}
\FALabel(2.34911,7.18253)[tl]{$q$}
\FAProp(20.,15.)(17.,10.)(0.,){/Straight}{-1}
\FALabel(17.6509,12.8175)[br]{$t$}
\FAProp(20.,5.)(17.,10.)(0.,){/Straight}{1}
\FALabel(19.3491,7.81747)[bl]{$t$}
\FAProp(3.,10.)(7.,10.)(0.,){/Cycles}{0}
\FALabel(5.,11.77)[b]{$g$}
\FAProp(17.,10.)(13.,10.)(0.,){/Cycles}{0}
\FALabel(15.,8.23)[t]{$g$}
\FAProp(7.,10.)(13.,10.)(0.8,){/GhostDash}{-1}
\FALabel(10.,6.53)[t]{$\eta$}
\FAProp(7.,10.)(13.,10.)(-0.8,){/GhostDash}{1}
\FALabel(10.,13.47)[b]{$\eta$}
\FAVert(3.,10.){0}
\FAVert(17.,10.){0}
\FAVert(7.,10.){0}
\FAVert(13.,10.){0}
\end{feynartspicture}
%
%
\begin{feynartspicture}(300,300)(1,1)
\FADiagram{}
\FAProp(0.,15.)(3.,10.)(0.,){/Straight}{1}
\FALabel(0.650886,12.1825)[tr]{$q$}
\FAProp(0.,5.)(3.,10.)(0.,){/Straight}{-1}
\FALabel(2.34911,7.18253)[tl]{$q$}
\FAProp(20.,15.)(17.,10.)(0.,){/Straight}{-1}
\FALabel(17.6509,12.8175)[br]{$t$}
\FAProp(20.,5.)(17.,10.)(0.,){/Straight}{1}
\FALabel(19.3491,7.81747)[bl]{$t$}
\FAProp(3.,10.)(7.,10.)(0.,){/Cycles}{0}
\FALabel(5.,11.77)[b]{$g$}
\FAProp(17.,10.)(13.,10.)(0.,){/Cycles}{0}
\FALabel(15.,8.23)[t]{$g$}
\FAProp(7.,10.)(13.,10.)(0.8,){/Cycles}{0}
\FALabel(10.,6.53)[t]{$g$}
\FAProp(7.,10.)(13.,10.)(-0.8,){/Cycles}{0}
\FALabel(10.,13.47)[b]{$g$}
\FAVert(3.,10.){0}
\FAVert(17.,10.){0}
\FAVert(7.,10.){0}
\FAVert(13.,10.){0}
\end{feynartspicture}
%
%
\begin{feynartspicture}(300,300)(1,1)
\FADiagram{}
\FAProp(0.,15.)(6.,10.)(0.,){/Straight}{1}
\FALabel(2.48771,11.7893)[tr]{$q$}
\FAProp(0.,5.)(6.,10.)(0.,){/Straight}{-1}
\FALabel(3.51229,6.78926)[tl]{$q$}
\FAProp(20.,15.)(14.,10.)(0.,){/Straight}{-1}
\FALabel(16.4877,13.2107)[br]{$t$}
\FAProp(20.,5.)(18.2,6.5)(0.,){/Straight}{1}
\FALabel(19.6123,6.46074)[bl]{$t$}
\FAProp(6.,10.)(14.,10.)(0.,){/Cycles}{0}
\FALabel(10.,8.93)[t]{$g$}
\FAProp(14.,10.)(15.8,8.5)(0.,){/Straight}{-1}
\FALabel(15.4123,9.96074)[bl]{$t$}
\FAProp(18.2,6.5)(15.8,8.5)(-0.8,){/Straight}{1}
\FALabel(15.6877,5.82926)[tr]{$t$}
\FAProp(18.2,6.5)(15.8,8.5)(0.8,){/Cycles}{0}
\FALabel(18.3123,9.17074)[bl]{$g$}
\FAVert(6.,10.){0}
\FAVert(14.,10.){0}
\FAVert(18.2,6.5){0}
\FAVert(15.8,8.5){0}
\end{feynartspicture}
%
%
\begin{feynartspicture}(300,300)(1,1)
\FADiagram{}
\FAProp(0.,15.)(6.,10.)(0.,){/Straight}{1}
\FALabel(2.48771,11.7893)[tr]{$q$}
\FAProp(0.,5.)(6.,10.)(0.,){/Straight}{-1}
\FALabel(3.51229,6.78926)[tl]{$q$}
\FAProp(20.,15.)(18.2,13.5)(0.,){/Straight}{-1}
\FALabel(18.5877,14.9607)[br]{$t$}
\FAProp(20.,5.)(14.,10.)(0.,){/Straight}{1}
\FALabel(17.5123,8.21074)[bl]{$t$}
\FAProp(6.,10.)(14.,10.)(0.,){/Cycles}{0}
\FALabel(10.,8.93)[t]{$g$}
\FAProp(14.,10.)(15.8,11.5)(0.,){/Straight}{1}
\FALabel(14.3877,11.4607)[br]{$t$}
\FAProp(18.2,13.5)(15.8,11.5)(-0.8,){/Straight}{-1}
\FALabel(18.3123,10.8293)[tl]{$t$}
\FAProp(18.2,13.5)(15.8,11.5)(0.8,){/Cycles}{0}
\FALabel(15.6877,14.1707)[br]{$g$}
\FAVert(6.,10.){0}
\FAVert(18.2,13.5){0}
\FAVert(14.,10.){0}
\FAVert(15.8,11.5){0}
\end{feynartspicture}

\caption{
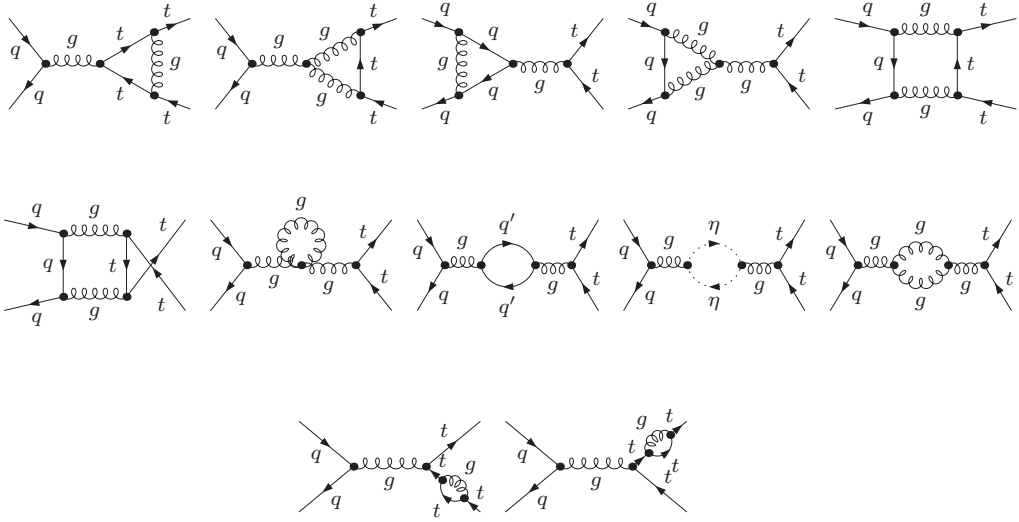
NLO QCD corrections to the process~(\ref{Eq:QQTTpro}). The ghosts are labelled by $\eta$ while $q'=u,d,c,s,t,b$.}
\label{Fig:QQTTloop}
}

\section{On-shell bubbles in  the spinor-helicity formalism}
\label{Sec:Spinors}

In this section we illustrate our method in the spinor-helicity formalism.
We compute the bubble coefficient of an amplitude  from the usual cut  integral using  
double cut spinor integration~\cite{Britto:2005ha, Britto:2006fc, Britto:2006sj, Britto:2007tt, Britto:2008vq, Mastrolia:2009dr}.

As discussed in Section~\ref{Sec:Wavefunction}, the spinor-helicity formalism requires a gauge choice of a null ``reference'' momentum for each gluon.  On-shell quantities are independent of this choice, but it plays a role in our method because of the off-shell continuation.   
Of course, the total bubble coefficient will be gauge-invariant.  However, the independence of gauge 
choice will not be apparent until all counterterm diagrams are considered, including the internal ones.

The cut integral involves the off-shell continuation of a  three-point  amplitude which  is a varying function of $q$, 
the reference momentum for the cut gluon. 
Furthermore, the off-shell current  $\widehat{\cal A}_{c c_{\rm ext}}$   (defined in Section~\ref{SSSec:BubDC})   depends on the reference 
momenta chosen for the external gluons.

In practice, one makes a convenient choice of reference momenta.  
It is important to be consistent in this choice throughout the entire calculation. 
The gauge choice of the reference momentum $q$ of the cut gluon cancels out immediately between the cut  and its counterterm.   In the following example, we
keep  $q$ general to show this cancellation, while making convenient choices for the reference momenta of external gluons. \\ 

The spinor-helicity formalism can accommodate massive particles as follows \cite{Kleiss:1985yh,Schwinn:2005pi}.  
For a particle of mass $m$ and momentum $p$, the massive spinors are given by
\bea
& & \ket{p} = \frac{(\slashed p+m)\sqket{\eta}}{\cb{p^\flat~ \eta}}, \qquad
\sqket{p} = \frac{(\slashed p+m)\ket{\eta}}{\vev{p^\flat~ \eta}},  \pagebreak[1]  
\\ &&
\bra{p} = \frac{\sqbra{\eta}(\slashed p+m)}{\cb{\eta ~p^\flat }}, \qquad
\sqbra{p} = \frac{\bra{\eta}(\slashed p+m)}{\vev{\eta ~p^\flat}},  \pagebreak[1]  
\eea
where $\eta$ is a fixed null momentum, and we define the null vector 
\bea
p^\flat = p - \frac{m^2}{2 p \cdot \eta} \eta.
\eea
For an antiparticle, the mass $m$ should be replaced by $-m$ everywhere.

It follows that massive spinor products obey the following (anti)symmetry relations:
\bea
\vev{ij}=-\vev{ji}, \qquad \cb{ij}=-\cb{ji}, \qquad \gb{ij}=\tgb{ji}.
\eea
 See Appendix~\ref{App:MS} for further details.

\medskip
As an example, we now focus on the process
\be
\bar t (k_1,c_1,h_1)\;  t (k_2,c_2,h_3)\; \to \;  g (k_3,c_3,h_3) \; g(k_4,c_4,h_4), 
\ee
where $h_i$ ($c_i$) is the helicity (color) of the $i^{th}$ particle.  We will compute the
coefficient of  $B_0(m_t^2;m_t,0)$ of  the leading-color part of the all-minus helicity 
amplitude.  Contributions to the on-shell bubble coefficients come from
the unrenormalized amplitude  $\mathcal{M}^{(1)}_{\bar t t \to gg} $ 
in the $k_1^2$ and $k_2^2$ channels  and from the counterterm diagrams.

\subsection{Contributions of the unrenormalized amplitude}
At one loop and in $D=4$ dimensions, the color decomposition of the unrenormalized amplitude 
reads as follows~\cite{Bern:1994fz,Badger:2011yu}:
\begin{eqnarray}
\mathcal{M}^{(1)}_{\bar t t \to gg} &=& \frac{2 g^4}{(4 \pi)^2}  \Bigg  [
 \sum_{(i,j) \in \{(3,4),(4,3) \}
}  N  \left( T^{c_i} T^{c_j} \right )^{c_1}_{c_2} A^{(1)}_{4;1}\left (1_{\bar t}^{h_1},2_{t}^{h_2},i_{g}^{h_i},j_{g}^{h_j}  \right )  
\nonumber \\ 
&& \qquad \qquad 
+ \mbox{tr}\left ( T^{c_3} T^{c_4}\right ) \delta^{c_1}_{c_2} A^{(1)}_{4;3}\left (1_{\bar t}^{h_1},2_{t}^{h_2},
3_{g}^{h_3} , 4_{g}^{h_4} \right )  \Bigg  ],
\label{Eq:MainSpin}
\end{eqnarray}
where $N$ is the number of colors.
The one-loop partial amplitudes $A^{(1)}_{4;1}$ are the leading-color ones, while 
$A^{(1)}_{4;3}$ is 
subleading.\footnote{The factor of $2$ is related to the different normalization of the color matrices 
used through the paper  and the ones in the color-ordered Feynman rules as in \cite{Dixon:1996wi}. } 

The external leg corrections of the anti-top are leading color, thus we  focus on the
leading-color partial amplitudes. We compute the bubble coefficient of 
$A^{(1)}_{4;1}\left (1_{\bar t}^{-},2_{t}^{-},3_{g}^{-},4_{g}^{-}  \right )$. The coefficient of
 $A^{(1)}_{4;1}\left (1_{\bar t}^{-},2_{t}^{-},4_{g}^{-},3_{g}^{-}  \right )$
can be obtained from the previous one via the substitution  $3 \leftrightarrow 4$.  The leading-color
partial amplitude can be written in terms of primitive amplitudes as~\cite{Bern:1994fz}
\begin{eqnarray}
A^{(1)}_{4;1}\left (1_{\bar t}^{-},2_{t}^{-},3_{g}^{-},4_{g}^{-}  \right ) &=& 
A^{L}\left (1_{\bar t}^{-},2_{t}^{-},3_{g}^{-},4_{g}^{-}  \right )  
-\frac{1}{N^2}A^{R}\left (1_{\bar t}^{-},2_{t}^{-},3_{g}^{-},4_{g}^{-}  \right )  \nonumber \\
&+& \frac{1}{N} 
\sum_q   A^{L,[1/2]}_{q}\left (1_{\bar t}^{-},2_{t}^{-},3_{g}^{-},4_{g}^{-}  \right ), 
\end{eqnarray}
where the sum runs over the quark species. 
The primitive  amplitudes contributing to the coefficient we are interested in are the left amplitude
$A^L$ and the right amplitude $A^R$.  They can be obtained starting from 
the color-ordered Feynman rules~\cite{Dixon:1996wi} and selecting the diagrams 
without a closed fermion loop, where the fermion flow  passes to the left  ($A^L$) or
the right ($A^R$) of the loop.   
\FIGURE[t]{
\unitlength=0.47bp%
%
%
%
\begin{feynartspicture}(300,300)(1,1)
\FADiagram{}
\FAProp(12.6,10.)(19.,10.)(0.,){/Cycles}{0}
\FAProp(19.,16.)(12.6,10.5)(0.,){/Straight}{1}
\FAProp(12.6,9.5)(19., 4.)(0.,){/Cycles}{0}
\FALabel(17.5,15.8)[b]{\scriptsize $2^-$}
\FALabel(19.,5.97)[b]{\scriptsize $4^-$}
\FALabel(19.,11.07)[b]{\scriptsize $3^-$}
\FAProp(0.,10.)(6.5,10.)(0.,){/Straight}{-1}
\FALabel(1.,11.07)[b]{\scriptsize $1^-$}
\FAProp(12.5,10.)(6.5,10.)(0.8,){/Straight}{1}
\FAProp(12.5,10.)(6.5,10.)(-0.8,){/Cycles}{0}
\FAVert(12.5,10.){-3}
\FAVert(6.5,10.){0}
\FAProp(9.25,15.)(9.25,5.)(0.,){/ScalarDash}{0}
\FALabel(9.3,0.)[b]{$\Delta_{2,k} A^{R}$}
\FAProp(5.7,8.)(8.3,6.5)(0.2,){/Straight}{-1}
\FALabel(7.,5.25)[b]{\scriptsize $\ell$}
\end{feynartspicture}
\hspace{0.61cm} 
\begin{feynartspicture}(300,300)(1,1)
\FADiagram{}
\FAProp(12.6,10.)(19.,10.)(0.,){/Cycles}{0}
\FAProp(19.,16.)(12.6,10.5)(0.,){/Straight}{1}
\FAProp(12.6,9.5)(19., 4.)(0.,){/Cycles}{0}
\FALabel(17.5,15.8)[b]{\scriptsize $2^-$}
\FALabel(19.,5.97)[b]{\scriptsize $4^-$}
\FALabel(19.,11.07)[b]{\scriptsize $3^-$}
\FAProp(0.,10.)(6.5,10.)(0.,){/Straight}{-1}
\FALabel(1.,11.07)[b]{\scriptsize $1^-$}
\FAProp(12.5,10.)(6.5,10.)(-0.8,){/Straight}{1}
\FAProp(12.5,10.)(6.5,10.)(0.8,){/Cycles}{0}
\FAProp(5.7,12.)(8.3,13.5)(-0.2,){/Straight}{-1}
\FALabel(7.,14.75)[t]{\scriptsize $\ell$}
\FAVert(12.5,10.){-3}
\FAVert(6.5,10.){0}
\FAProp(9.25,15.)(9.25,5.)(0.,){/ScalarDash}{0}
\FALabel(9.3,0.)[b]{$\Delta_{2,k} A^{L}$}
\end{feynartspicture}
%
%
%
%
%
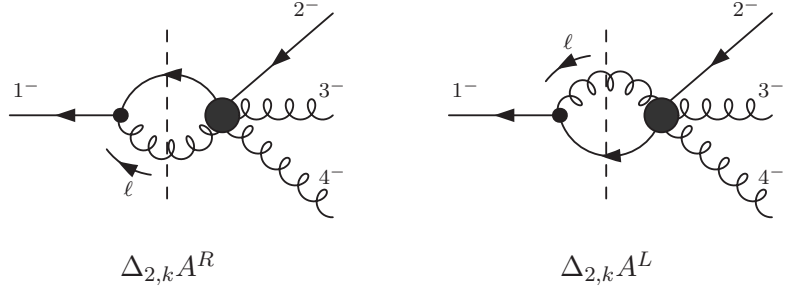
\caption{$k_1^2$-channel cut of the  right  and left primitive amplitudes.}
\label{Fig:DoubleCutSpin}
} 
\subsubsection{The cut in the $k^2_1$ channel}
 We perform  the shift setting $\bar k = k_4$, i.e. 
\be
k_1 \to  \hat k_1 = k_1 + \xi k_4, \qquad  k_4 \to \hat k_4 = (1+\xi) k_4,
\qquad \gamma=\Spapb{4}{1}{4}.
\label{Eq:bla}
\ee
Note that these shifts have the same sign, since $k_1$ is incoming while $k_4$ is outgoing, in contrast to our all-outgoing convention of Section \ref{Sec:Wavefunction}.

The double cut of the right  amplitude $A^R$, Figure~\ref{Fig:DoubleCutSpin},  reads as follows:
\be
\Delta_{2,-\hat k_1} A^R\left (1_{\bar t}^{-},2_{t}^{-},3_{g}^{-},4_{g}^{-}  \right ) 
=  \int d \mu_{2,-\hat k_1} \; I^R.
\label{Eq:CutSpinR}
\ee
The integrand $I^R$ is written in terms  of  color-ordered tree-level helicity amplitudes as
\be
I^R =
 \sum_{h_t, h_g = \pm}  
A\left (1_{\bar t}, -(\ell+\hat k_1)^{ -h_t}_{t},  -\ell^{-h_g}_g \right ) 
A\left ( (\ell+\hat k_1)^{ h_t}_{\bar t}, 2^{-}_{t},  3^-_g,  4^-_g,\ell^{h_g}_g  \right ).
\ee
The necessary helicity amplitudes are collected in Appendix~\ref{App:Tree}.

After the spinor integration and the $\xi$ expansion, the bubble coefficient 
$b^R_{B_0(m_t^2 +   \xi \gamma ;m_t,0)}$  can be written in 
a relatively compact form:
\bea
 b^R_{B_0(m_t^2 +   \xi \gamma ;m_t,0)}  &=&
 \frac{1}{\xi}
\frac{2  m_t^3 \vev{34} \cb{12} }{  \cb{34}\gb{4|1|4}^2}
+
\frac{  m_t^2  \vev{34} (m_t [2q] \vev{q 4} \cb{4 1} + [1 4 \rangle \gb{q|1|2} \cb{4 q})
}{\cb{34}\gb{4|1|4}^2\gb{q|1|q}}
\nonumber \\
 &-&  \frac{m_t}{\cb{3 4} \gb{4|1|4}^2} \bigg [
m_t \langle 4 1] \cb{4 2} \vev{3 4}  + m_t^2 \langle 3 2] \langle4 1] + m_t \gb{4|1|2} [1 3 \rangle 
\nonumber \\
&& \qquad \qquad 
-2 \cb{1 2} \vev{3 4} \gb{4|1|4} 
 \bigg ].
\label{Eq:CoeffRSpin}
\eea
The  dependence  on $q$ comes from the off-shell three-point
function. See equations (\ref{eq:3ptqm}) and (\ref{eq:3ptqp}). This dependence will drop out in the sum with the counterterm diagram.   

The double cut of the left amplitude $A^L$ reads as follows (cf. Figure~\ref{Fig:DoubleCutSpin}):
\be
\Delta_{2,-\hat k_1} A^L\left (1_{\bar t}^{-},2_{t}^{-},3_{g}^{-},4_{g}^{-}  \right ) 
=  \int d \mu_{2,-\hat k_1} I^L,
\label{Eq:CutSpinL}
\ee
with the integrand $I^L$ defined as 
\be
I^L =
 \sum_{h_t, h_g = \pm}  
A\left (1_{\bar t},-\ell^{-h_g}_g, -(\ell+\hat k_1)^{ -h_t}_{t} \right ) 
A\left ( (\ell+\hat k_1)^{ h_t}_{\bar t}, \ell^{h_g}_g,  2^{-}_{t},  3^-_g,  4^-_g \right ).
\ee
 By considering the properties of the color-ordered Feynman
rules, the $\bar q  g q gg$ amplitude can be expressed as a
linear combination of  $\bar q qggg$  amplitudes, as depicted in  Figure~\ref{Fig:TreeSpin}.
Therefore $I^L$ can be written as $I^L=I^R+I^l$, where  $I^{l}$ is given by 
\begin{eqnarray}
I^{l} &=&  \sum_{h_t, h_g = \pm}    A\left (1_{\bar t}, -(\ell+\hat k_1)^{ -h_t}_{t},  -\ell^{-h_g}_g \right )  \bigg [
A\left ((\ell+\hat k_1)^{ h_t}_{\bar t}, 2^{-}_{t},  3^-_g, \ell^{h_g}_g,    4^-_g   \right ) \nonumber \\
&&  + A\left ((\ell+\hat k_1)^{ h_t}_{\bar t}, 2^{-}_{t},  \ell^{h_g}_g,  3^-_g,  4^-_g   \right ) \
\bigg ].
\end{eqnarray}
It turns out that the bubble coefficient from $I^l$  vanishes.
Therefore, the bubble coefficient of the left amplitude,  $b^L_{B_0(m_t^2 +   \xi\gamma ;m_t,0)}$ 
is equal to the one of the right amplitude: 
\be
b^L_{B_0(m_t^2 +   \xi \gamma ;m_t,0)} = b^R_{B_0(m_t^2 +   \xi \gamma ;m_t,0)}.
\label{Eq:CoeffLSpin}
\ee

The final contribution to the bubble coefficient of the leading color part of the amplitude~(\ref{Eq:MainSpin}) is obtained 
using the results~(\ref{Eq:CoeffRSpin})  and~(\ref{Eq:CoeffLSpin}) and expanding $B_0(m^2_t+\xi\gamma;m_t,0)$
around $\xi =0$. 
The outcome is
\begin{eqnarray}
&& 2 g^2  \Bigg \{ \frac{g^2C_F}{8 \pi^2} \left( T^{c_3} T^{c_4} \right )^{c_1}_{c_2} \Bigg [ 
b^R_{B_0(m_t^2 +   \xi \gamma ;m_t,0)} \;  B_0(m_t^2;m_t,0) \nonumber \\
&&- \frac{2  m_t^3 \vev{34} \cb{12} }{  \cb{34}\gb{4|1|4}} \; B'_0(m_t^2;m_t,0)  \Bigg ]  + \left ( 3 \leftrightarrow 4 \right ) \Bigg \}.
\label{Eq:BCoeffSpin}
\end{eqnarray}
\FIGURE[t]{
\unitlength=0.33bp%
\begin{feynartspicture}(300,300)(1,1)
\FADiagram{}
\FAProp(8.,10.)(8.,18.)(0.,){/Straight}{-1}
\FALabel(9.4,18)[]{\scriptsize $2$}
\FAProp(8.,10.)(15.6085,12.4721)(0.,){/Cycles}{0}
\FALabel(15.,13.8)[]{\scriptsize $3$}
\FAProp(8.,10.)(0.3915,12.4721)(0.,){/Cycles}{0}
\FALabel(1.,13.8)[]{\scriptsize $\ell$}
\FAProp(8.,10.)(12.7023,3.52786)(0.,){/Cycles}{0}
\FALabel(14.,3.5)[]{\scriptsize $4$}
\FAProp(8.,10.)(3.2977,3.52786)(0.,){/Straight}{1}
\FALabel(0.7,3.5)[]{\scriptsize $\ell +k_1$}
\FAVert(8.,10.){-3}
\FALabel(19.0,10)[]{=}
\end{feynartspicture}
\hspace{0.5cm}
\begin{feynartspicture}(300,300)(1,1)
\FADiagram{}
\FAProp(8.,10.)(8.,18.)(0.,){/Cycles}{0}
\FALabel(9.4,18)[]{\scriptsize $3$}
\FAProp(8.,10.)(15.6085,12.4721)(0.,){/Cycles}{0}
\FALabel(15.,13.8)[]{\scriptsize $4$}
\FAProp(8.,10.)(0.3915,12.4721)(0.,){/Straight}{-1}
\FALabel(1.,13.8)[]{\scriptsize $2$}
\FAProp(8.,10.)(12.7023,3.52786)(0.,){/Cycles}{0}
\FALabel(14.,3.5)[]{\scriptsize $\ell$}
\FAProp(8.,10.)(3.2977,3.52786)(0.,){/Straight}{1}
\FALabel(0.7,3.5)[]{\scriptsize $\ell +k_1$}
\FAVert(8.,10.){-3}
\FALabel(-2.70,10)[]{--}
\end{feynartspicture}
\begin{feynartspicture}(300,300)(1,1)
\FADiagram{}
\FAProp(8.,10.)(8.,18.)(0.,){/Cycles}{0}
\FALabel(9.4,18)[]{\scriptsize $3$}
\FAProp(8.,10.)(15.6085,12.4721)(0.,){/Cycles}{0}
\FALabel(15.,13.8)[]{\scriptsize $\ell$}
\FAProp(8.,10.)(0.3915,12.4721)(0.,){/Straight}{-1}
\FALabel(1.,13.8)[]{\scriptsize $2$}
\FAProp(8.,10.)(12.7023,3.52786)(0.,){/Cycles}{0}
\FALabel(14.,3.5)[]{\scriptsize $4$}
\FAProp(8.,10.)(3.2977,3.52786)(0.,){/Straight}{1}
\FALabel(0.7,3.5)[]{\scriptsize $\ell +k_1$}
\FAVert(8.,10.){-3}
\FALabel(-2.70,10)[]{--}
\end{feynartspicture}
\begin{feynartspicture}(300,300)(1,1)
\FADiagram{}
\FAProp(8.,10.)(8.,18.)(0.,){/Cycles}{0}
\FALabel(9.4,18)[]{\scriptsize $\ell$}
\FAProp(8.,10.)(15.6085,12.4721)(0.,){/Cycles}{0}
\FALabel(15.,13.8)[]{\scriptsize $3$}
\FAProp(8.,10.)(0.3915,12.4721)(0.,){/Straight}{-1}
\FALabel(1.,13.8)[]{\scriptsize $2$}
\FAProp(8.,10.)(12.7023,3.52786)(0.,){/Cycles}{0}
\FALabel(14.,3.5)[]{\scriptsize $4$}
\FAProp(8.,10.)(3.2977,3.52786)(0.,){/Straight}{1}
\FALabel(0.7,3.5)[]{\scriptsize $\ell +k_1$}
\FAVert(8.,10.){-3}
\FALabel(-2.70,10)[]{--}
\end{feynartspicture}
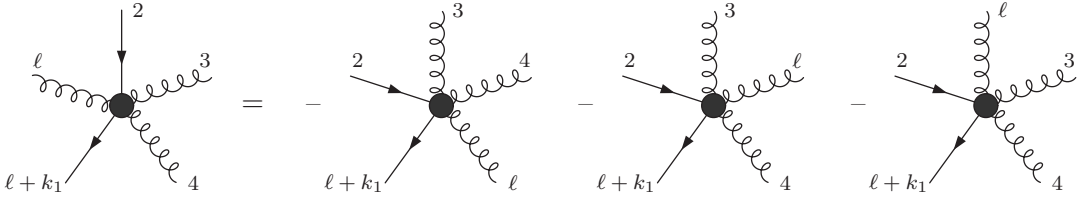
\caption{Pictorial representation of the  relation between  the $\bar q  g q gg$ and  the 
$\bar q qggg$ color-ordered tree-level amplitudes.}
\label{Fig:TreeSpin}
} 
\subsubsection{The cut in the $k^2_2$ channel}
In this case, the symmetry of the amplitude allows us to write down the result as a simple relabeling of the contribution from the $k^2_1$ channel.  
It suffices to exchange $1\leftrightarrow 2, 3\leftrightarrow 4$, and $m_t \to -m_t$.  The result is 
\begin{eqnarray}
&& 2 g^2  \Bigg \{ \frac{g^2C_F}{8 \pi^2} \left( T^{c_3} T^{c_4} \right )^{c_1}_{c_2} \Bigg [ 
b^R_{B_0(m_t^2 +   \xi \gamma ;m_t,0)} \big |_{1\leftrightarrow 2,  m_t \to -m_t} \;  B_0(m_t^2;m_t,0) \nonumber \\
&&- \frac{2  m_t^3 \vev{34} \cb{12} }{  \cb{34}\gb{4|2|4}} \; B'_0(m_t^2;m_t,0)  \Bigg ]  + \left ( 3 \leftrightarrow 4 \right ) \Bigg \}.
\label{Eq:BCoeffSpin2}
\end{eqnarray}

\subsection{Contributions of the counterterm diagrams}
The counterterm diagrams can be color decomposed as follows:
\be
\mathcal{M}^{\rm ct}_{\bar t t \to gg} = 2 g^2 \bigg [ 
\left( T^{c_3} T^{c_4} \right )^{c_1}_{c_2} A^{\rm ct}\left (1_{\bar t}^{-},2_{t}^{-},3_{g}^{-},4_{g}^{-}  \right ) +
\left( T^{c_4} T^{c_3} \right )^{c_1}_{c_2} A^{\rm ct}\left (1_{\bar t}^{-},2_{t}^{-},
4_{g}^{-} , 3_{g}^{-} \right )
\bigg ].
\label{Eq:CTSpin}
\ee
The color-ordered counterterm diagram  $A^{\rm ct}$  can be computed using the 
color-ordered Feynman rules after performing the shift~(\ref{Eq:bla}).  We will focus on
$ A^{\rm ct}\left (1_{\bar t}^{-},2_{t}^{-},3_{g}^{-},4_{g}^{-}  \right )$,
since $A^{\rm ct}\left (1_{\bar t}^{-},2_{t}^{-}, 4_{g}^{-} , 3_{g}^{-} \right )$ can be 
obtained via the substitution $3 \leftrightarrow 4$.  The corresponding 
 color ordered counterterm  diagrams are depicted in Figure~\ref{Fig:ctSpin}.

The  diagram $D_a$ in  Figure~\ref{Fig:ctSpin}  is related to  
the tree-level amplitudes where  $1_{\bar t}$ is continued
off-shell.  These amplitudes can be derived by a recursion relation as described in Appendix~\ref{App:Tree}. 
Using a decomposition similar to the one in eq.~(\ref{Eq:CTexpanded0}), the 
counterterm diagram $D_a$ can be written as
$D_a  =  A^{m} + A^{\psi}+A^{k}$,  whose bubble terms can be read
from eq.~(\ref{Eq:CTexpanded2}),  (\ref{Eq:CTexpanded1}), and~(\ref{Eq:CTexpanded3}).
Substituting the expansion of $J$ from  (\ref{Eq:BCFWcurr}) into these expressions,  we find
\begin{eqnarray}
A^m &=& -\frac{g^2 C_F}{8 \pi^2} \Bigg \{  
   \frac{1}{\xi  }   B_0(m_t^2;m_t,0)  \;
\frac{2m_t^3\vev{34}\cb{12}}{\gb{4|1|4}^2 \cb{34}}
 -  B_0(m_t^2;m_t,0)  \;
 \frac{m_t^2 \cb{1|3|2} }{\gb{4|1|4}\cb{34}^2}  
 \nonumber \\ && \qquad \qquad 
+ B_0(m_t^2;m_t,0) \;  
\frac{m_t \vev{34}(\cb{14}\gb{4|1|4}\cb{32}+m_t\tgb{14}\cb{43}\cb{42})}{\gb{4|1|4}^2\cb{34}^2}
 \nonumber \\ && \qquad \qquad 
- B_0(m_t^2;m_t,0) \;  \frac{m_t \cb{1|4|3|2}}{2\gb{4|1|4}\cb{34}^2}
\Bigg\}, 
\nonumber  \pagebreak[1]   \\ 
A^{\psi}  &=& -\frac{g^2 C_F}{8 \pi^2}  B'_0(m_t^2;m_t,0) \;
\Bigg \{  
  \frac{ 2 m_t^3 \vev{34}\cb{12}}{\gb{4|1|4}\cb{34}}
\Bigg\},  \nonumber  \pagebreak[1]   \\
A^{k} &=&\frac{g^2 C_F}{8 \pi^2}  B_0(m_t^2;m_t,0)
\Bigg \{   
 \;  \frac{m_t^2 \vev{34}(m_t\cb{q2}\cb{14}\vev{4q}+\gb{q|1|2}\tgb{14}\cb{q4})}{\gb{4|1|4}^2\gb{q|1|q}\cb{34}}
-\frac{m_t\cb{12}\vev{34}}{\gb{4|1|4}\cb{34}}
\Bigg \}.     \nonumber
\end{eqnarray}
We have kept the full $q$-dependence above, in order to demonstrate that it cancels exactly between the bubble computed from the cut 
and the counterterm.  In practice, we might set $q=k_4$ throughout, in which case the $q$-dependent  terms simply vanish.

The external leg counterterm diagram $D_b$ in Figure~\ref{Fig:ctSpin} can be obtained from $D_a$ by 
performing the substitutions $1\leftrightarrow 2, \; 3\leftrightarrow 4$, and $m_t \to -m_t$.

Finally, there are contributions to the coefficient of  $B_0(m_t^2;m_t,0)$ coming from all other self-energy counterterms.  In this case, there is only one, namely diagram $D_c$.  The counterterm separates the scattering process into two off-shell currents.  For the example at hand, the currents are simply the cubic interactions, and the Feynman rules give the bubble coefficient  
\begin{eqnarray}
D_c &=& \frac{g^2 C_F}{8 \pi^2}\frac{m_t }{\Spapb{4}{1}{4}^2 [34]^2} \; \Bigg [
\Big (2 m_t^2 - \Spapb{4}{1}{4} \Big ) \Big ( [ 1 4\rangle [34]  \langle 3 2] + [13] \vev{4 3} [42] \Big  ) \nonumber \\
&&  \qquad \qquad +
2 m_t  \Spapb{4}{1}{4}  \Big ( [1 4 \rangle [4 2] - [1 3] \langle 3 2] \Big ) 
\Bigg ].
\label{jjcorr}
\end{eqnarray}

The bubble contribution of the counterterm diagrams, eq.~(\ref{Eq:CTSpin}), is then
\be
 2 g^2 \bigg [ 
\left( T^{c_3} T^{c_4} \right )^{c_1}_{c_2}  \big ( D_a + D_b + D_c \big ) +  \big ( 3 \leftrightarrow 4 \big )
\bigg ].
\label{Eq:CTSpinF}
\ee

\FIGURE[t]{
\unitlength=0.42bp
%
%
%
%
\begin{feynartspicture}(300,300)(1,1)
\FADiagram{}
\FAProp(0.,15.)(10.,10.)(0.,){/Straight}{1}
\FALabel(2.48771,12.3)[tr]{\scriptsize$2$}
\FAProp(0.,5.)(3.,6.5)(0.,){/Straight}{-1}
\FALabel(2.01229,5.2)[tl]{\scriptsize$1$}
\FAProp(20.,15.)(10.,10.)(0.,){/Cycles}{0}
\FALabel(16.4877,13.7)[br]{\scriptsize$3$}
\FAProp(20.,5.)(10.,10.)(0.,){/Cycles}{0}
\FALabel(17.5123,7.5)[bl]{\scriptsize$4$}
\FAProp(3.,6.5)(10.,10.)(0.,){/Straight}{-1}
\FAVert(10.,10.){-3}
\FAVert(4.,7.0){2}
\FALabel(10.0,0.)[]{$D_a$}
\end{feynartspicture}
\begin{feynartspicture}(300,300)(1,1)
\FADiagram{}
\FAProp(0.,15.)(4.,13.0)(0.,){/Straight}{1}
\FALabel(0.987714,13.0393)[tr]{\scriptsize$2$}
\FAProp(0.,5.)(10.,10.)(0.,){/Straight}{-1}
\FALabel(3.51229,6.2)[tl]{\scriptsize$1$}
\FAProp(20.,15.)(10.,10.)(0.,){/Cycles}{0}
\FALabel(16.4877,13.7)[br]{\scriptsize$3$}
\FAProp(20.,5.)(10.,10.)(0.,){/Cycles}{0}
\FALabel(17.5123,7.5)[bl]{\scriptsize$4$}
\FAProp(4.,13.0)(10.,10.)(0.,){/Straight}{1}
\FAVert(10.,10.){-3}
\FAVert(4.,13.0){2}
\FALabel(10.0,0.)[]{$D_b$}
\end{feynartspicture}
\begin{feynartspicture}(300,300)(1,1)
\FADiagram{}
\FAProp(0.,15.)(10.,14.)(0.,){/Straight}{1}
\FALabel(4.84577,13.4377)[t]{\scriptsize$2$}
\FAProp(0.,5.)(10.,6.)(0.,){/Straight}{-1}
\FALabel(5.15423,4.43769)[t]{\scriptsize$1$}
\FAProp(20.,15.)(10.,14.)(0.,){/Cycles}{0}
\FALabel(14.8458,15.5623)[b]{\scriptsize$3$}
\FAProp(20.,5.)(10.,6.)(0.,){/Cycles}{0}
\FALabel(15.1542,6.56231)[b]{\scriptsize$4$}
\FAProp(10.,10.)(10.,14.)(0.,){/Straight}{-1}
\FAProp(10.,10.)(10.,6.)(0.,){/Straight}{1}
\FAVert(10.,14.){0}
\FAVert(10.,6.){0}
\FAVert(10.,10.){2}
\FALabel(10.0,0.)[]{$D_c$}
\end{feynartspicture}

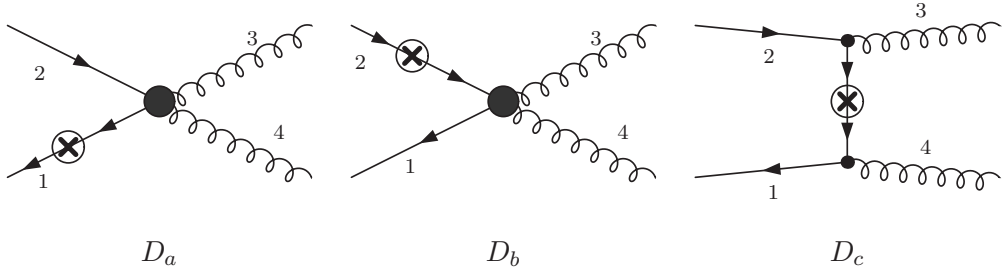
\caption{Counterterm diagrams contributing to the coefficient of the on-shell bubble $B_0(m^2_t; m_t; 0)$. }
\label{Fig:ctSpin}
}

\subsection{Total bubble coefficient}

Here, we assemble all the contributions to the coefficient of $B_0(m_t^2;m_t,0)$. The bubble coefficient coming from the cuts and counterterms  is
obtained summing the contributions~(\ref{Eq:BCoeffSpin}), (\ref{Eq:BCoeffSpin2}), and~(\ref{Eq:CTSpinF}).  After the cancellation of divergent and 
gauge-dependent pieces between the cut-diagram and counterterm contributions, and then recovering the on-shell limit, we find the following result:
\begin{eqnarray}
&& \frac{g^4 C_F}{4 \pi^2}\frac{m_t \left( T^{c_3} T^{c_4} \right )^{c_1}_{c_2}  }{\Spapb{4}{1}{4}^2 [34]^2} \; \Bigg [
-2m_t^2\cb{13}\vev{34}\cb{42}+2m\gb{4|1|4} \Big(\tgb{13}\cb{32}+\tgb{14}\cb{42}\Big)
\nonumber \\ && 
-m_t s_{34}\Big(\tgb{13}\cb{32}+\cb{14}\gb{42}\Big)
-\vev{34} \gb{4|1|4}\Big(\cb{13}\cb{42}-\cb{14}\cb{32} \Big)
\Bigg ] + \big (3 \leftrightarrow 4 \big ).  
\end{eqnarray}

To compute the bubble coefficient for the entire amplitude, one would need the contributions from all remaining helicity amplitudes.  It is easy to translate the results above to any other helicity amplitude whose gluons have equal helicities, either by overall parity conjugation of the amplitude, or by conjugating the individual massive spinors.  The conventions for massive spinors (see Appendix \ref{App:MS}) allow their individual conjugation except in spinor strings where they appear on both ends; these strings can be split using Schouten identities.  Thus the remaining nontrivial computation for the leading-color  part would be any single configuration with opposite helicities for the  two gluons.

There are also coefficients multiplying the differentiated bubble, $B'_0(m_t^2;m_t,0)$, arising from the various counterterms via $\delta Z_{\psi}$.  They are proportional to the tree-level amplitude.  The counterterm for external wavefunction renormalization cancels exactly with the cut diagram in this coefficient, as shown in Section \ref{Sec:Wavefunction}, but other counterterms need to be treated separately.

\section{Summary}
\label{Sec:Summary}

We close this article with a brief review of our procedure.

\begin{enumerate}

\item 
For each massive external fermion, with momentum denoted by $k$, choose the shift direction $\bar{k}$.  Define the off-shell continuation 
$\hat k = k + \xi \bar{k}$,
where $\xi$ parametrizes the shift, and choose another external leg to absorb the shift in order to conserve momentum, as in equation (\ref{shift}).  

\item
Compute the tree-level inputs for the cuts in the shifted momentum channels.    The cuts may diverge as $1/\xi$.  The expressions are needed through $\ord(\xi^0)$.  The expansion might be done by a contour integral, or by generating the $\ord(\xi^{-1})$ and $\ord(\xi^0)$ terms separately.

\item
The cut integrals have now been regulated, and the coefficients of master integrals $A_0$ and $B_0$ can be found in the usual way.  Keep only the $\ord(\xi^0)$ term.  The $1/\xi$-divergent term is ignored, unless it is desired for a consistency check.  It will be proportional to the tree amplitude, as shown in equations (\ref{Eq:DCexpanded}) and (\ref{Eq:SCexpanded}).  Because of the off-shell continuation, the cut depends on the gauge choice made for the cut gluon.

\item 
Compute the counterterms from tree-level currents.  The counterterms for external leg corrections are computed at $\ord(\xi^0)$.  The divergent part cancels the divergence of the loop exactly and may likewise be ignored.  Similarly, the coefficient of $B'_0$ will cancel and may be discarded.  In the spinor-helicity method, one must include the term ${\cal M}_k$ from equation (\ref{Eq:CTexpanded3}), making the same gauge choice for the cut gluon as in the cut integral.

All other counterterms for internal legs and vertices are finite and can be computed from unshifted expressions.  
Considered separately, the counterterms depend on the gauge choices made for external legs.  Gauge invariance is restored in the sum. It is the internal counterterms that provide the surviving contributions to the differentiated bubble $B'_0$, via the renormalization constant $\delta Z_{\psi}$.

\end{enumerate}

\section*{Acknowledgments}
We are grateful to Leandro Almeida for collaborating  in the  early
stages of this project. We  thank Zvi Bern and
David Kosower for insightful comments and feedback on a draft of this manuscript, and Alexander Ochirov for corrections to the first version.
We acknowledge Simon Badger and  Harald Ita for useful discussions.  This research was supported in part by the National Science Foundation under Grant No. NSF PHY05-51164;
we thank  the KITP  for its hospitality.   R.B. is supported by the Agence Nationale de la Recherche under grant ANR-09-CEXC-009-01. E.M. is supported by the European Research Council under Advanced Investigator Grant ERC-AdG-228301.

\appendix

\section{Master integrals, double and single cuts}
\label{App:MI}

The master integrals that pertain to the wavefunction renormalization diagram are the scalar bubble,
\bean
B_0(k^2;m,0) \equiv {(2\pi\mu)^{4-D}  \over i \pi^2} \int{d^{D}\ell } {1\over \ell^2 ((k-\ell)^2-m^2)},
\eean
and the tadpole with massive propagator,
\bean
A_0(m) \equiv {(2\pi\mu)^{4-D} \over i \pi^2} \int {d^{D}\ell } {1 \over(k-\ell)^2-m^2}.
\eean
These integrals are dimensionally regularized by taking the dimensionality $D=4-2\eps$. The double cut of the bubble and the single cut of the tadpole are obtained by substituting the propagators with the corresponding 
delta functions:
\begin{eqnarray}
\Delta_{2,k} B_0(k^2;m,0) &\equiv& {(2\pi\mu)^{4-D}  \over i \pi^2} \int{d^{D}\ell }\;  \delta(\ell^2) \;  \delta\left((k-\ell)^2-m^2 \right  ), \nonumber \\
\Delta_{1,k} A_0(m) &\equiv& {(2\pi\mu)^{4-D}  \over i \pi^2} \int{d^{D}\ell } \;  \delta \left ((k-\ell)^2-m^2 \right  ).
\end{eqnarray}
The rule for computing the double cut from the left and right tree amplitudes is
\bea
\Delta_{2,k} \mathcal{M}  = \int d \mu_{2,k} \;  \mathcal{M}_L(\ell) \mathcal{M}_R(\ell), 
~~~ d \mu_{2,k} \equiv  - \frac{1}{(2 \pi)^4} \; d^D \ell \; \delta(\ell^2) \;  \delta \left ((k-\ell)^2-m^2 \right ).
\label{2cut-lips}
\eea
The rule for computing the single cut from tree amplitudes is 
\bea
\Delta_{1,k} \mathcal{M}  = \int d \mu_{1,k} \; \mathcal{M}_T(\ell),  
\qquad d \mu_{1,k} \equiv  \frac{i}{(2 \pi)^4}  \;  d^D \ell \;  \delta \left ((k-\ell)^2-m^2 \right ).
\label{1cut-lips}
\eea
For convenience, we list the reduction of several double-cut  integrals, 
\begin{eqnarray}
\int d \mu_{2,k}  &=& -\frac{i }{16 \pi^2}\; \Delta_{2,k} B_0(k^2;m,0), \nonumber \\
\int d \mu_{2,k}  (R\cdot \ell)  &=&-\frac{i }{16 \pi^2}  \frac{(k^2-m^2)}{2 k^2}  (R\cdot k) \; \Delta_{2,k} B_0(k^2;m,0), \nonumber \\
\int d \mu_{2,k}   \frac{(R\cdot \ell)}{(\ell + p)^2 -m^2}&=& -\frac{i }{32 \pi^2} \frac{k^2(p\cdot R) - (k\cdot p) (k\cdot R)   }{k^2 p^2 - (k\cdot p)^2} \; \Delta_{2,k} B_0(k^2;m,0), 
\label{Eq:ResB0}
\end{eqnarray}
and single-cut integrals,
\begin{eqnarray}
\int d \mu_{1,k} \frac{1}{(\ell+p)^2} &=& 0 , \nonumber \\
\int d \mu_{1,k} \frac{(R\cdot \ell)}{(\ell+p)^2}  &=&- \frac{1 }{32 \pi^2}  \frac{R\cdot (k+p)}{(k+p)^2} \; \Delta_{1,k} A_0(m).
\label{Eq:ResA0}
\end{eqnarray}
Here, $R$ and $p$ are any 4-vectors.

\section{Massive spinors}
\label{App:MS}

We use the formalism of massive spinors developed by Kleiss and Stirling \cite{Kleiss:1985yh}.  

For assistance in calculations, we have used the specific representation of massive spinors in terms of an arbitrary null ``reference'' momentum $\eta$, which is the same for all particles, as follows \cite{Schwinn:2005pi}.    For an on-shell particle of mass $m$ and momentum $p$, define
\bea
p^\flat = p - \frac{m^2}{2 p \cdot \eta} \eta,
\eea
which is another null vector.

Then the massive spinors are defined as 
\bea
\ket{p} = \frac{(\slashed p+m)\sqket{\eta}}{\cb{p^\flat~ \eta}}, \qquad
\sqket{p} = \frac{(\slashed p+m)\ket{\eta}}{\vev{p^\flat~ \eta}},
\eea
whose conjugates are
\bea
\bra{p} = \frac{\sqbra{\eta}(\slashed p+m)}{\cb{\eta ~p^\flat }}, \qquad
\sqbra{p} = \frac{\bra{\eta}(\slashed p+m)}{\vev{\eta ~p^\flat}}.
\eea

For an antiparticle, the mass $m$ should be replaced by $-m$ everywhere.
These formulas are smooth in the massless limit.

We have also used the \verb+Mathematica+ package \verb+S@M+~\cite{Maitre:2007jq}
to help with spinor manipulations and numerical evaluation.

\subsection{Some identities for massive spinors}

Among the identities obeyed by the massive spinors, we find the following particularly useful in our computations.

The contractions of spinors of the same chirality are antisymmetric as usual, while the mixed spinor product is symmetric:
\bea
\vev{ij}=-\vev{ji}, \qquad \cb{ij}=-\cb{ji}, \qquad \gb{ij}=\tgb{ji}.
\eea

The spinor representation of a momentum vector can be replaced by the massive spinors using
\bea
\slashed p &=& \ket{p}\sqbra{p}+\sqket{p}\bra{p}-m   \qquad \mbox{for a particle,} \\
\slashed p &=& \ket{p}\sqbra{p}+\sqket{p}\bra{p}+m   \qquad \mbox{for an antiparticle}.
\eea
The slash is implicit inside spinor products. Note that
$
\gb{i|j|i} = 2 p_i \cdot p_j.
$

Finally, the  Schouten identity is still valid for any four spinors of the same type,
\be
 \vev{ij}\vev{kl}+\vev{ik}\vev{lj}+\vev{il}\vev{jk}=0, \qquad
 \cb{ij}\cb{kl}+\cb{ik}\cb{lj}+\cb{il}\cb{jk}=0,
\label{MassiveSchouten}
\ee
while of course the three-spinor uncontracted versions remain valid for any massless spinors $a,b,c$,
\be
\ket{a}\vev{bc}+\ket{b}\vev{ca}+\ket{c}\vev{ab} =0, \qquad
 \sqket{a}\cb{bc}+\sqket{b}\cb{ca}+\sqket{c}\cb{ab}=0.
\ee

\subsection{Tree-level formulas}
\label{App:Tree}
Here we list the tree amplitudes needed for the cut in the example of Section~\ref{Sec:Spinors}.  They are adapted from \cite{Hall:2007mz}.\footnote{Equivalent formulas were derived earlier in \cite{Ozeren:2006ft,Schwinn:2006ca,Schwinn:2007ee}, also by on-shell recursion relations.} 
The three- and five-point amplitudes, which are ingredients in the cut computation, are given in a form that allows straightforward parity conjugation of the fermions. 

The convention we follow here is that gluon momenta are directed outward, while quark and 
anti-quark momenta are directed inward.  The mass of the quark is $m$.

\paragraph{Three-point amplitudes.} 
The relevant three-point amplitudes, valid also in the off-shell continuation, are
\bea
A(1_{\bar q}^-,2_q^-,3_g^-) &=& 
\sqbra{1}\left(\frac{\sqket{q}\bra{3}+\ket{3}\sqbra{q}}{\cb{q3}}\right)\sqket{2}, \label{eq:3ptqm}
\\
A(1_{\bar q}^-,2_q^-,3_g^+) &=& 
\sqbra{1}\left(\frac{\ket{q}\sqbra{3}+\sqket{3}\bra{q}}{\vev{q3}}\right)\sqket{2}. \label{eq:3ptqp}
\eea
The quantity inside parentheses is the polarization vector, which depends on the null reference momentum $q$. On-shell, the amplitude is independent of $q$.

\paragraph{Four-point amplitudes.} 
The tree-level amplitude corresponding to the one-loop amplitude of interest is
\bea
A(1_{\bar q}^-,2_q^-,3_g^-,4_g^-) &=&
\frac{m\vev{34}\cb{12}}{(2 p_2 \cdot p_3)\cb{34}}.
\eea

\paragraph{Five-point amplitudes.} 
The amplitudes entering the double cuts~(\ref{Eq:CutSpinR}) and~(\ref{Eq:CutSpinL}) are
\bea
A(1_{\bar q}^+,2_q^-,3_g^-,4_g^-,5_g^-)&=&
\frac{m\vev{3| p_{45}|1|5}(\gb{14}\cb{32}-\gb{13}\cb{42})}{(2 p_1 \cdot p_5)(2 p_2 \cdot p_3) \cb{34}^2 \cb{45}},   \pagebreak[1]  \\[1.6ex]
A(1_{\bar q}^+,2_q^-,3_g^-,4_g^-,5_g^+)&=&
\frac{\tgb{5|1| p_{34}|2|3} (\cb{25}\vev{1| p_{34}|2|3}-\tgb{2| p_{34}|d_{51}|3}\gb{15})}{(2 p_1 \cdot p_5)(2 p_2 \cdot p_3) \cb{34} \cb{45}\vev{5|p_{34}|2|3}}\nonumber \\
&& - \frac{m\vev{43}^3(\vev{1| p_{34}|5}\gb{3| p_{45}|2}-\vev{1| p_{45}|3}\gb{5| p_{34}|2})}{(p_1+p_2)^4 \vev{45}\vev{35}\vev{5| p_{34}|2|3}},  \pagebreak[1]  \\[1.6ex]
A(1_{\bar q}^+,2_q^-,3_g^-,4_g^+,5_g^-)&=&  \frac{m\vev{35}^3(\vev{1| p_{34}|5}\gb{3| p_{45}|2}-\vev{1| p_{45}|3}\gb{5| p_{34}|2})}{(p_1+p_2)^4 \vev{34}\vev{45}\vev{5| p_{34}|2|3}}\nonumber \\
&& -\frac{\tgb{4|1|5}\tgb{4|2|3}}{(2 p_1 \cdot p_5)(2 p_2 \cdot p_3) \cb{34} \cb{45}\vev{5| p_{34}|2|3}}
\Big (m\gb{14}\vev{53}\cb{42} \nonumber \\
&&+\vev{15}\cb{42}\tgb{4|2|3}  -\gb{14}\gb{32}\tgb{4|1|5} \Big ), \pagebreak[1]  \\[1.6ex]
A(1_{\bar q}^+,2_q^-,3_g^+,4_g^-,5_g^-)&=&
- \frac{\tgb{3|2| p_{45}|1|5}(\gb{13}\gb{5|1| p_{45}|2}+\vev{1| p_{45}| d_{32}|5}\cb{32})}{(2 p_1 \cdot p_5)(2 p_2 \cdot p_3) \cb{34} \cb{45}\vev{5|1| p_{45}|3}} \nonumber \\
&& + \frac{m\vev{45}^3 \left(\vev{1| p_{34}|5}\gb{3| p_{45}|2}-\vev{1| p_{45}|3}\gb{5| p_{34}|2}\right)}{(p_1+p_2)^4 \vev{34}\vev{35}\vev{5|1| p_{45}|3}},  \pagebreak[1] 
\eea
where $p_{ij} \equiv p_i +p_j$ and $d_{ij} \equiv p_i -p_j$.
Similar amplitudes with the opposite helicity choice for $1_{\bar q}$  are given by exchanging angle brackets $\ket{1},\bra{1}$ with square brackets $\sqket{1},\sqbra{1}$.  (The quark $2_q$ can be conjugated similarly, but we do not need that in the present paper.)  This type of conjugation is valid as long as the derivation of the tree amplitudes did not apply any identities such as the Schouten identities, (\ref{MassiveSchouten}), which does not respect conjugation of a single massive spinor if it is contracted with another.

\paragraph{Off-shell four-point currents.} 
For the counterterm, we use a more general formula for the tree-level four-point current.
This expression is valid when  the  particle  $1_{\bar q}$ is continued off shell and   its external spinor is stripped off. Because it is off shell, it depends on the reference momenta $q_3$ and $q_4$ of gluons $3$ and $4$, respectively.
The full expression is
\bea
J_{q_3,q_4}(1_{\bar q}^-,2_q^-,3_g^-,4_g^-) &=& 
 \frac{1}{\cb{q_3 3}\cb{q_4 4}}
\Bigg[\frac{1}{\Spapb{3}{2}{3}}
\bigg ( \cb{1 q_4} \gb{4|1|q_3} \gb{32} - \tgb{14} \tgb{q_4|2|3} \cb{q_3 2} \nonumber \\
&&    - m \cb{1 q_4}\vev{43}\cb{q_3 2}-m\tgb{14}\cb{q_4 q_3}\gb{32} \bigg ) 
\nonumber \\
&& +\frac{1}{\cb{34}}
\bigg ( \cb{1 q_4}\gb{42}\cb{q_3 4}  + \tgb{14} \cb{q_4 2} \cb{q_3 4} \nonumber \\
&&+ \cb{1 q_3}\gb{32}\cb{q_4 3}   + \tgb{13}\cb{q_3 2}\cb{q_4 3}
+\frac{\cb{1|d_{34}|2}\cb{q_3 q_4}}{2}
\bigg )
\Bigg].
\eea

For the example discussed in Section \ref{Sec:Spinors}, we need the current for the process where only $k_1$ is off shell, and we choose $q_3=k_4, q_4=k_3$ throughout.  This current, with the spinor $\sqbra{1}$ stripped off, is given by   
\bea
J &=& \frac{p_{34}\ket{2|3}\cb{3 2}
+m\sqket{3}\vev{43}\cb{42}
 + (m^2-p_{1}^2)\sqket{3}\tgb{23}}{2 p_2 \cdot p_3 \cb{34}^2}
 -\frac{p_{34}\sqket{2}}{2\cb{34}^2}.
\label{Eq:BCFWcurr}
\eea

\bibliographystyle{JHEP}
\bibliography{references}

\end{document}